\documentclass[journal,twocolumn]{IEEEtran}
\PassOptionsToPackage{table,dvipsnames}{xcolor}
\usepackage{latexsym}
\usepackage{epsfig}
\usepackage{amsfonts,amsmath,amssymb,amstext}
\usepackage{pifont}
\usepackage{graphicx}
\usepackage{psfrag}
\usepackage{dsfont}
\usepackage{cite}
\usepackage{float}

\DeclareMathAlphabet\mathbfcal{OMS}{cmsy}{b}{n}
\usepackage{hyperref}
\usepackage{multirow}
\usepackage{xcolor}
\usepackage[most]{tcolorbox}
\usepackage{tikz}
\usetikzlibrary{positioning}
\usepackage{array}
\usepackage{enumitem}
\usepackage{ifthen,calc,graphicx,pgfplots}
\usepackage[caption=false]{subfig}
\usepackage{makecell}
\usepackage{orcidlink}
\pgfplotsset{compat=1.15}
\usetikzlibrary{plotmarks}
\usetikzlibrary{arrows}

\input{acronyms}

\usepackage{amsmath,amssymb,amsthm,amsbsy}
\usepackage{xifthen}
\usepackage{mathtools}
\usepackage{enumerate}
\usepackage{microtype}
\usepackage{xspace}
\usepackage{bm}
\usepackage[T1]{fontenc}
\usepackage{bbm}

\newcommand{\mbs}[1]{\pmb{#1}}
\newcommand{\vect}[1]{{\lowercase{\mbs{#1}}}}
\newcommand{\mat}[1]{{\uppercase{\mbs{#1}}}}

\newcommand{\rvMat}[1]{{\pmb{#1}}} 

\renewcommand{\H}{{\scriptscriptstyle\mathsf{H}}}

\newcommand{\hv}{\vect{h}}

\newcommand{\nv}{\vect{n}}

\newcommand{\sv}{\vect{s}}

\newcommand{\uv}{\vect{u}}
\newcommand{\vv}{\vect{v}}
\newcommand{\wv}{\vect{w}}
\newcommand{\xv}{\vect{x}}
\newcommand{\yv}{\vect{y}}

\newcommand{\muv}{\vect{\mu}}

\newcommand{\Am}{\mat{a}}
\newcommand{\Bm}{\mat{b}}
\newcommand{\Cm}{\mat{c}}
\newcommand{\Dm}{\mat{d}}

\newcommand{\Gm}{\mat{g}}
\newcommand{\Hm}{\mat{h}}

\newcommand{\Qm}{\mat{q}}

\newcommand{\Sm}{\mat{s}}

\newcommand{\Um}{\mat{u}}
\newcommand{\Vm}{\mat{V}}
\newcommand{\Wm}{\mat{w}}
\newcommand{\Xm}{\mat{x}}
\newcommand{\Ym}{\mat{y}}
\newcommand{\Zm}{\mat{z}}

\newcommand{\Sigmam}{\pmb{\Sigma}}

\newcommand{\Cc}{{\mathcal C}}

\newcommand{\Nc}{{\mathcal N}}

\newcommand{\Xc}{{\mathcal X}}

\newcommand{\CC}{{\mathbb C}}

\newcommand{\GG}{{\mathbb G}}

\newcommand{\NN}{{\mathbb N}}

\newcommand{\RR}{{\mathbb R}}

\newcommand{\Id}{\mat{\mathit{I}}} 

\newcommand{\CN}[1][]{\ifthenelse{\isempty{#1}}{\mathcal{N}_{\mathbb{C}}}{\mathcal{N}_{\mathbb{C}}\left(#1\right)}}

\renewcommand{\P}[1][]{\ifthenelse{\isempty{#1}}{\mathbb{P}}{\mathbb{P}\left[#1\right]}}

\newcommand{\E}[1][]{\ifthenelse{\isempty{#1}}{\mathbb{E}}{\mathbb{E}\left[#1\right]}}

\newcommand{\I}[1][]{\ifthenelse{\isempty{#1}}{\mathbb{I}}{\mathbb{I}\left\{#1\right\}}}

\renewcommand{\det}[1][]{\ifthenelse{\isempty{#1}}{\mathrm{det}}{\mathrm{det}\left(#1\right)}}

\newcommand{\trace}[1][]{\ifthenelse{\isempty{#1}}{\mathrm{tr}}{\mathrm{tr}\left(#1\right)}}

\newcommand{\rank}[1][]{\ifthenelse{\isempty{#1}}{\mathrm{rank}}{\mathrm{rank}\left(#1\right)}}

\newcommand{\diag}[1][]{\ifthenelse{\isempty{#1}}{\mathrm{diag}}{\text{diag}\left(#1\right)}}

\newcommand{\Cov}[1][]{\ifthenelse{\isempty{#1}}{\mathsf{Cov}}{\mathsf{Cov}\left[#1\right]}}

\newcommand{\Span}[1][]{\ifthenelse{\isempty{#1}}{\mathsf{Span}}{\mathsf{Span}\left(#1\right)}}

\DeclarePairedDelimiter\abs{\lvert}{\rvert}
\DeclarePairedDelimiter\norm{\lVert}{\rVert}
\DeclarePairedDelimiter\normf{\lVert}{\rVert_{\text{F}}}

\renewcommand{\Re}[1][]{\ifthenelse{\isempty{#1}}{\operatorname{Re}}{\operatorname{Re}\left(#1\right)}}
\renewcommand{\Im}[1][]{\ifthenelse{\isempty{#1}}{\operatorname{Im}}{\operatorname{Im}\left(#1\right)}}

\newcommand{\ind}[1]{{\mathbbm{1}\!\left\{#1\right\}}}

\newcommand{\Nt}{M}
\newcommand{\Nr}{N}

\newcommand{\defeq}{=}

\def\X{{\bf X}}

\def\Gras{\mathbb{G}(M,{\mathbb{C}}^{T})}
\def\St{\mathbb{S}_t(M,{\mathbb{C}}^{T})}

\newcommand{\cmark}{\ding{51}}

\newcommand{\note}[1]{{\color{blue} #1}}
\renewcommand{\note}[1]{}

\begin{document}
\title{Noncoherent MIMO Communications: \\ Theoretical Foundation, Design Approaches, \\ and Future Challenges}

\author{Khac-Hoang Ngo\textsuperscript{\orcidlink{0000-0003-2047-6957}}, \emph{Member, IEEE,} Diego Cuevas\textsuperscript{\orcidlink{0000-0001-6101-1182}}, \\ Ruben de Miguel Gil\textsuperscript{\orcidlink{0009-0009-2838-4775}}, \emph{Student Member, IEEE,} Victor Monzon Baeza\textsuperscript{\orcidlink{0000-0002-0035-3944}}, \emph{Senior Member, IEEE,} \\ Ana Garcia Armada\textsuperscript{\orcidlink{0000-0002-8495-6151}}, \emph{Fellow, IEEE,} and Ignacio Santamaria\textsuperscript{\orcidlink{0000-0003-0040-7436}}, \emph{Senior Member, IEEE}
\thanks{Khac-Hoang Ngo is with the Department of Electrical Engineering, Linköping University, 58183 Linköping, Sweden~(e-mail: {khac-hoang.ngo@liu.se}).}
\thanks{Diego Cuevas and Ignacio Santamaria are with the Department of Communications
Engineering, Universidad de Cantabria, 39005 Santander, Spain (e-mails:
\{diego.cuevas, i.santamaria\}@unican.es).}
\thanks{Ruben de Miguel Gil and Ana Garcia Armada are with the Signal Theory and Communications Department, Universidad Carlos III de Madrid, 28911
Legan\'es, Madrid, Spain (e-mails: \{rumiguel, anagar\}@ing.uc3m.es).}
\thanks{Victor Monzon Baeza is with Universitat Oberta de Catalunya (UOC), Barcelona, Spain (e-mail: vmonzon@uoc.edu).}
\thanks{The work of Khac-Hoang Ngo was supported by the Excellence Center at Linköping–Lund in Information Technology (ELLIIT). The work of Diego Cuevas and Ignacio Santamaria was supported by grant PID2022-137099NB-C43 (MADDIE) funded by
MICIU/AEI /10.13039/501100011033 and FEDER, UE. The work of Diego Cuevas was also partly supported under grant FPU20/03563 funded by MIU, Spain. The work of Ana Garcia Armada and Ruben de Miguel Gil was supported by project SOFIA-AIR (PID2023-147305OB-C31)(MICIU /10.13039/501100011033 / AEI / EFDR, UE) and Horizon Europe project MiFuture under grant agreement number 101119643. 
}
}

\thispagestyle{empty}
\maketitle

\begin{abstract} 
    Noncoherent communication is a promising paradigm for future wireless systems where acquiring accurate channel state information (CSI) is challenging or infeasible. It provides methods to bypass the need for explicit channel estimation in practical scenarios such as high-mobility networks, massive distributed antenna arrays, energy-constrained Internet-of-Things devices, and unstructured propagation environments.
    This survey provides a comprehensive overview of noncoherent communication strategies in multiple-input multiple-output (MIMO) systems, focusing on recent advances since the early 2000s. We classify noncoherent communication schemes into three main approaches where CSI-free signal recovery is based on subspace detection (i.e., Grassmannian signaling), differential detection, and energy detection, respectively. For each approach, we review the theoretical foundation and design methodologies. We also provide comparative insights into their suitability across different channel models and system constraints, highlighting application scenarios where noncoherent methods offer performance and scalability advantages over traditional coherent communication. Furthermore, we discuss practical considerations of noncoherent communication, including compatibility with orthogonal frequency division multiplexing (OFDM), resilience to hardware impairments, and scalability with the number of users. Finally, we provide an outlook on future challenges and research directions in designing robust and efficient noncoherent systems for next-generation wireless networks.
\end{abstract}

\begin{IEEEkeywords}
    Noncoherent communication, MIMO, channel state information, Grassmannian signaling, differential detection, energy-based detection.
\end{IEEEkeywords}

\section{Introduction \note{--Hoang}}
\label{sec:intro}

Wireless communication has long been the backbone of modern connectivity. The evolution of wireless communication technology through its generations has enabled global mobile networks, satellite communications, and broadband wireless access. A central challenge in wireless communication is dealing with channel fading, i.e., the variation of channel impulse responses\footnote{Throughout the paper, we also refer to channel impulse responses as channel state/coefficients/vector/matrix, or simply as the channel.} due to multipath propagation, mobility, and environmental factors. Fading causes variations in signal strength and phase over time, frequency, and space, necessitating sophisticated techniques to mitigate its impact. Traditionally, reliable data transmission over fading channels has relied on \textit{coherent communication}, where the receiver exploits accurate knowledge of the channel state to optimally decode the transmitted signals. The transmitter can also exploit this knowledge, if available, to adapt its transmission strategy. For a \gls{MIMO} channel, coherent communication efficiently exploits the spatial dimensions by decomposing the channel into multiple parallel, non-interfering \gls{SISO} channels and multiplexing independent data onto these channels~\cite{Goldsmith03_capacityMIMO}. 

Coherent communication 
relies on the availability of \gls{CSI} at the receiver and/or transmitter. This is typically achieved through the transmission of known reference symbols, so-called pilots, from the transmitter to the receiver to estimate the channel, and through channel state feedback from the receiver to the transmitter. The channel estimate is then used for communication when the channel has not changed significantly. 
The variation of wireless channels 
requires frequent transmissions of pilot symbols to maintain accurate channel estimates. However, acquiring and maintaining \gls{CSI} is difficult or even infeasible 
in many emerging application scenarios, such as massive-scale \gls{IoT} networks, \gls{HRLLC}, 
and high-mobility vehicular and aerial networks. 
We list some examples of such scenarios in the following.
\begin{itemize}
    \item \textit{High mobility and rapidly-varying channels:} In highly mobile environments, such as in vehicular and drone-based 
    communications, 
    the channel remains stable for a small amount of time and frequency samples, requiring frequent and costly \gls{CSI} updates.  As a rule of thumb, an $M$-antenna transmitter needs to send at least $M$ pilot symbols within each channel coherence interval, within which the channel remains constant, for the receiver to determine $M$ channel vectors corresponding to the transmit antennas~\cite{Hassibi2003howmuchtraining}. If the coherence interval lasts for $T$ channel uses, there remain $T - M$ channel uses for data transmission before the channel needs to be estimated again. If the channel changes rapidly, $T$ is short, and the fraction of pilot transmission becomes disproportionate to data transmission, especially if $M$ is large.

    
    \item \textit{Distributed and massive antenna arrays:} 
    Orthogonal pilot-based \gls{CSI} acquisition in massive or gigantic \gls{MIMO} systems introduces a considerable overhead \cite{PopovskiTSP16}. 
    The pilot length can be limited by letting the users, having a small number of antennas, transmit the pilot signal~\cite{Marzetta16_massiveMIMO}. However, the complexity of channel estimation still scales with the number of antennas. Furthermore, for the estimated uplink channel to be used in the downlink, the channel has to be reciprocal, and the base station antennas have to be phase synchronized. In the case of distributed antennas with independent local oscillators, maintaining reciprocity and phase alignment requires frequent transmission of reference signals between the antennas~\cite{Dorfler14_synchronization,Larsson24_synchrony}. This makes conventional \gls{CSI}-based approaches inefficient in certain scenarios.

    \item \emph{Massive connectivity:} In the uplink of a multi-user system, pilot sequences are assigned per user and orthogonally across users. If the total number of users is larger than the coherence interval (but probably only a random number of users are active at a time), pre-assigning mutually orthogonal pilot sequences to every user present in the system is impossible.  
    This is the case for \gls{mMTC}. Reusing pilot sequences among interfering users leads to pilot contamination~\cite{Dey21_pilotContamination}, preventing the separation of the channels of these users. One can consider nonorthogonal pilots, but accurate \gls{CSI} acquisition is still challenging.
        
    \item \emph{Energy-constrained devices:} In ultra-low-power applications, such as ambient \gls{IoT} with batteryless and backscatter devices~\cite{Butt24_ambientIoT}, transmitting pilot signals for channel estimation can be infeasible due to the strict energy limitations.

    \item \emph{Harsh and unstructured propagation environments}: In particular communication scenarios, such as in deep space~\cite{Xu19_Aeronautical}, underwater~\cite{Lidstrom25}, and channels with nonlinearities~\cite{Ngo21_GGM}, the channel exhibits unpredictable and non-stationary behavior, making channel estimation challenging and coherent detection unreliable.
\end{itemize}
    
Given these challenges, \emph{noncoherent communication} has emerged as a compelling alternative. In noncoherent communication, the signal transmission and reception are performed without instantaneous \gls{CSI}. Noncoherent communication can be designed with only statistical \gls{CSI}, i.e., the channel distribution, or even without this knowledge. 
By eliminating the need for explicit \gls{CSI}, noncoherent schemes offer enhanced robustness and reduced complexity. It has also been shown that noncoherent detection outperforms coherent detection, with the cost of pilot transmissions taken into account, in terms of ergodic capacity~\cite{ZhengTse2002Grassman}, finite-blocklength achievable rate~\cite{Ostman2019pilotVSnoncoherent}, and error exponent~\cite{Chowdhury15}. These advantages make noncoherent communication highly attractive for next-generation wireless networks where efficiency and scalability are critical. 

This survey provides a comprehensive review of recent advances in noncoherent MIMO communications regarding fundamental understanding,  system design, practical considerations, as well as a discussion on future challenges.

\subsection{Noncoherent Communication Approaches}
\label{sec:intro:approaches}
We focus on noncoherent schemes where no explicit channel estimation is performed. Under this condition, the design of noncoherent communications relies on identifying components of the transmitted signal that are invariant under the impact of fading, and thus can be detected at the \gls{CSI}-free receiver. 
Along this line, we next briefly describe three main approaches to noncoherent communications with multiple antennas. 
\begin{itemize}
    \item \textit{Schemes based on subspace detection:} Consider a transmitter sending a $T \times M$ symbol matrix over $M$ antennas and $T$ channel uses of a coherence interval. With low noise, the column space of this signal matrix can be reliably detected by a noncoherent receiver. This is because the unknown channel matrix only scales and rotates the basis of the transmitted signal matrix without changing its column space. Therefore, information carried in this subspace is preserved. This leads to a design approach of mapping information bits onto the column space of the signal matrix. This column space belongs to the Grassmannian manifold $\Gras$, i.e., the space of $M$-dimensional subspaces of $\CC^T$. Therefore, this approach is called \textit{Grassmannian signaling}. 

    \item \textit{Schemes based on energy detection:}
    A noncoherent receiver can detect the energy of a transmitted symbol by using the squared norm of the received signal as its statistics. This detection is reliable if the receiver has a large number of antennas for which channel hardening occurs. This gives rise to \textit{energy-based schemes} where information is conveyed on the amplitude of the transmitted signal.
    
    \item \textit{Schemes based on differential detection:}
    When the channel varies slightly and continuously between channel uses/blocks, the noncoherent receiver can reliably detect transitions of the transmitted signal between channel uses/blocks by comparing the corresponding received signals. This leads to \textit{differential-detection-based schemes} where information bits are mapped to the rotation between consecutively transmitted symbols. For a single-antenna transmitter, this rotation is between the symbol phase, as in~\gls{DPSK}. For an $M$-antenna transmitter, the rotation is represented by an $M\times M$ unitary matrix. 
    
\end{itemize}

\subsection{Paper Contributions}
In this survey, we review the development of noncoherent communication with a focus on schemes that follow the three mentioned approaches, exploit multiple antennas, and have been developed since the early 2000s. Our main contributions are summarized as follows:  
\begin{itemize}
    \item We present the application scenarios of noncoherent communication in the context of \gls{MIMO} fading channels. These scenarios include conventional \gls{MIMO} vs. massive/gigantic \gls{MIMO}, and block fading vs. continuously varying fading. 

    \item We provide an overview of the three design approaches for noncoherent communication, namely, Grassmannian signaling, energy-based schemes, and differential-detection-based schemes. We give a comparative discussion on the three approaches, in terms of the scenario they are most suitable for. 
    
    \item For each approach, we provide a structured review of the theoretical foundation and existing designs for both the single-user and multi-user scenarios. The review highlights theoretical benchmarks for the achievable performance of noncoherent communication, and a trade-off between performance and computational complexity of practical designs.
    
    \item We analyze the practical considerations and inherent limitations associated with the three approaches. We systematically examine key factors such as compatibility with \gls{OFDM} architectures, sensitivity to \gls{DC} components, robustness to \glspl{HWI}, spectral efficiency constraints, and scalability with respect to the number of users. This analysis offers critical insights into the trade-offs and design considerations that must be addressed when deploying noncoherent schemes in practical wireless communication systems.
        
    \item We put forth an outlook on the use of noncoherent communication in emerging applications and open challenges that need to be resolved for noncoherent communication to be adopted in next-generation wireless systems.
\end{itemize}

\subsection{Related Work}
The development of noncoherent communications has been reviewed in several existing survey and tutorial papers. In~\cite{roger2014noncoherent}, the authors presented the design principle of Grassmannian constellations and evaluated their performance in two practical scenarios, namely, a MIMO system with antenna correlation and a coordinated multi-point system with unbalanced
transmission points.  
Also focusing on Grassmannian signaling, \cite{Gohary2019noncoherentMIMOsignaling} discussed design approaches and practical implementation challenges. These works neither provided a comprehensive overview nor covered differential detection and energy-based schemes as in our paper.
A broad survey of sixty years of coherent versus noncoherent
trade-offs (from the late 1960s to the late 2010s) was presented in~\cite{xu2019sixty}. While comprehensive in its historical scope, \cite{xu2019sixty} primarily emphasized differential-detection-based schemes, and did not cover many important contributions to block transmission techniques.

Other studies highlight noncoherent communication as an enabler for emerging scenarios and applications. For instance,
\cite{Nawaz21} explored the integration of noncoherent detection with backscatter communication for massive connectivity. 
Meanwhile, \cite{Schwarz23multiple} discussed the potential of Grassmannian codebooks to cope with challenging channel characteristics toward \gls{6G} of wireless networks via \gls{CSI} quantization and feedback, and \gls{CSI}-free communication. Furthermore,~\cite[Sec.~IV]{Kaddoum16} reviews noncoherent analog and digital modulation schemes, but focuses on chaos-based communication systems. 

In Table~\ref{tab:surveys}, we summarize the scope of existing survey and tutorial papers in comparison to our paper. 
\begin{table*}[t!]
    \renewcommand{\arraystretch}{1.1}
    \caption{A comparison of existing surveys/tutorials on noncoherent communications and our contribution 
    }
    \centering
    \begin{tabular}{|m{2cm}|m{.7cm}|m{7cm}| >{\centering\arraybackslash}m{1.6cm} | >{\centering\arraybackslash}m{1.6cm} | >{\centering\arraybackslash}m{1.8cm} |} 
    \hline
    \textbf{Paper} & \textbf{Year} & \textbf{Scope} & \textbf{Grassmannian schemes} & \textbf{Energy-based schemes} & \textbf{Differential detection-based schemes}  \\
    \hline \hline
    Roger \textit{et al.}~\cite{roger2014noncoherent} & 2014 & Performance of Grassmannian signaling in practical scenarios & \cmark & & \cmark \\\hline
    Kaddoum~\cite{Kaddoum16} & 2016 & Noncoherent modulation schemes for chaos-based communication systems &  & & \cmark  \\ \hline 
    Gohary \textit{et al.}~\cite{Gohary2019noncoherentMIMOsignaling} & 2019 & Design approaches and practical implementation challenges of Grassmannian signaling & \cmark &   & \\ \hline
    Xu \textit{et al.}~\cite{xu2019sixty} & 2019 & Comprehensive survey of the coherent versus noncoherent trade-off &  & & \cmark \\ \hline
    Nawaz \textit{et al.}~\cite{Nawaz21} & 2021 & Noncoherent and backscatter communication for massive connectivity & \cmark & & \cmark  \\ \hline
    Schwarz \& Pratschner~\cite{Schwarz23multiple} & 2023 & Grassmannian codebook for quantized \gls{CSI} feedback and \gls{CSI}-free communication & \cmark & &  
    \\ \hline \hline
    Ours & 2025 & Theoretical foundation, design approaches, and future challenges of noncoherent MIMO communication & \cmark & \cmark & \cmark \\ \hline
    \end{tabular}
    \label{tab:surveys}
\end{table*}

\subsection{Paper Outline and Acronyms}
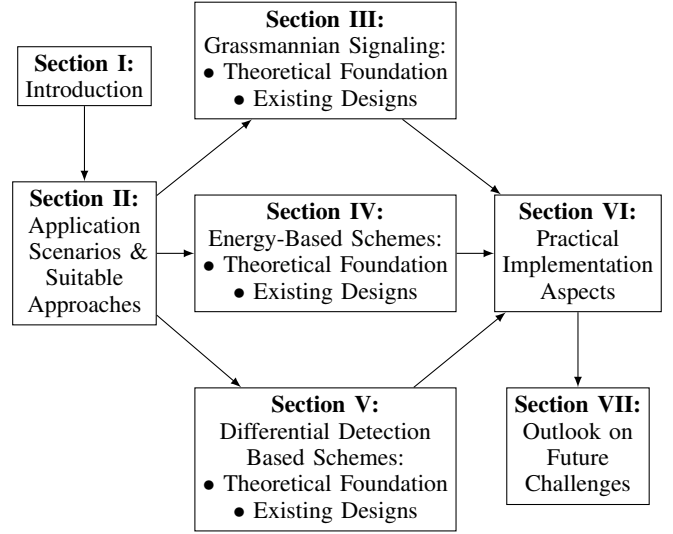
\begin{figure}[t!]
    \centering
    \begin{tikzpicture} \small
        \node[rectangle,draw=black,align=center] at (0,0) (model) {\bf Section~\ref{sec:model}: \\ Application \\\ Scenarios \& \\ Suitable \\ Approaches};

        \node[rectangle,draw=black,align=center,above=1cm of model] (intro) {\bf Section~\ref{sec:intro}: \\ Introduction};
        \node[rectangle,draw=black,align=center,right=.5cm of model] (energy) {\bf Section~\ref{sec:energy-based}: \\ Energy-Based Schemes: \\ $\bullet$ Theoretical Foundation \\  $\bullet$ Existing Designs};

        \node[rectangle,draw=black,align=center,above=1cm of energy] (grassmannian) {\bf Section~\ref{sec:Grassmannian}: \\ Grassmannian Signaling: \\ 
        $\bullet$ Theoretical Foundation \\ 
        $\bullet$ Existing Designs};

        \node[rectangle,draw=black,align=center,below=1cm of energy] (differential) {\bf Section~\ref{sec:differential}: \\ Differential Detection \\ Based Schemes: \\
        $\bullet$ Theoretical Foundation \\ $\bullet$ Existing Designs};

        \node[rectangle,draw=black,align=center,right=.5cm of energy] (practical) {\bf Section~\ref{sec:practical}: \\ Practical \\ Implementation \\ Aspects};
        \node[rectangle,draw=black,align=center,below=1cm of practical] (outlook) {\bf Section~\ref{sec:future_challenges}: \\ Outlook on \\ Future \\ Challenges};

        \draw[-latex] (intro) -- (model);
        
        \draw[-latex] (model) -- (grassmannian);
        \draw[-latex] (model) -- (energy);
        \draw[-latex] (model) -- (differential);
        \draw[-latex] (grassmannian) -- (practical);
        \draw[-latex] (energy) -- (practical);
        \draw[-latex] (differential) -- (practical);
        \draw[-latex] (practical) -- (outlook);
    \end{tikzpicture}
    \caption{The overall layout of the survey.}
    \label{fig:layout}
    \vspace{-.4cm}
\end{figure}
We depict the structure of the paper in Fig.~\ref{fig:layout}. The remainder of the paper is organized as follows. In Section~\ref{sec:model}, we present an overview of channel models and scenarios, the three main noncoherent design approaches, and a discussion about which scenario best fits each approach. Next, we review in detail the theoretical foundation and existing designs for each approach: Grassmannian signaling in Section~\ref{sec:Grassmannian}, energy-based schemes in Section~\ref{sec:energy-based}, and differential-detection-based schemes in Section~\ref{sec:differential}. Practical aspects and limitations of the three approaches are analyzed and discussed in Section~\ref{sec:practical}. In Section~\ref{sec:future_challenges}, we provide an outlook and put forth future challenges in the development of noncoherent schemes for emerging communication systems. We conclude the survey in Section~\ref{sec:conclusion}. In the appendices, we give a brief tutorial on packing in Riemannian manifolds.

Table~\ref{tab:acronym} summarizes the acronyms used in the paper. 
\begin{table}[t!]
    \small
    \centering
    \caption{List of Acronyms}
    \label{tab:acronym}
    \begin{tabular}{l l}
         5G     & the Fifth Generation \\
         6G     & the Sixth Generation \\
         ADC    & Analog-to-Digital Converter \\
         AI     & Artificial Intelligence \\
         APSK   & Amplitude and Phase Shift Keying \\
         ASK    & Amplitude Shift Keying \\
         AWGN   & Additive White Gaussian Noise \\
         BICM   & Bit-Interleaved Coded Modulation \\
         BC     & Broadcast Channel \\
         BER    & Bit Error Rate \\
         CFO    & Carrier Frequency Offset \\
         CRLB   & Cramér-Rao Lower Bound \\
         CSI    & Channel State Information \\
         DC     & Direct Current \\
         DBPSK  & Differential Binary Phase Shift Keying \\
         DFDD   & Decision-Feedback Differential Detection \\
         DPSK   & Differential Phase Shift Keying \\
         DQPSK  & Differential Quadrature Phase Shift Keying \\
         DoF    & Degree of Freedom \\
         DUSTM  & Differential Unitary Space-Time Modulation \\
         GLRT   & Generalized Likelihood Ratio Test \\ 
         HWI    & Hardware Impairment \\
         HRLLC  & Hyper-Reliable Low-Latency Communications \\
         i.i.d. & independent and identically distributed \\
         IM     & Index Modulation \\
         IoT    & Internet of Things \\
         IIoT   & Industrial Internet of Things \\
         IQI    & I/Q Imbalance \\
         ISAC   & Integrated Sensing and Communication \\
         KL     & Kullback–Leibler \\
         LNA    & Low-Noise Amplifier \\
         LTE    & Long Term Evolution \\
         LPD    & Low Probability of Detection\\
         OFDM   & Orthogonal Frequency Division Multiplexing \\
         OTFS   & Orthogonal Time Frequency Space \\
         MAC    & Multiple-Access Channel \\
         MIMO   & Multiple-Input Multiple-Output \\
         mMTC   & massive Machine-Type Communication \\
         mmWave & Milimeter Wave \\
         ML     & Maximum Likelihood \\
         MSDD   & Multiple-Symbol Differential Detection \\
         PAM    & Pulse Amplitude Modulation \\
         PEP    & Pairwise Error Probability \\
         PMF    & Probability Mass Function \\
         PSK    & Phase Shift Keying \\
         QEC    & Quantum Error Correction \\
         RIS    & Reconfigurable Intelligent Surface\\
         SER    & Symbol Error Rate \\
         SIMO   & Single-Input Multiple-Output \\
         SISO   & Single-Input Single-Output \\
         SNR    & Signal-to-Noise Ratio \\
         STBC    & Space-Time Block Code \\
         THz    & Terahertz \\
         UMA    & Unsourced Multiple Access \\
         URLLC  & Ultra-Reliable Low-Latency Communication \\
         USTM   & Unitary Space-Time Modulation \\
    \end{tabular}
\end{table}

\section{Application Scenarios and \\ Suitable Noncoherent Schemes \note{- Diego}} \label{sec:model}

In this section, we provide an overview of various channel models and scenarios in which noncoherent communications are advantageous. We also discuss the noncoherent schemes best suited to each scenario.


\subsection{Application Scenarios}

\subsubsection{Conventional MIMO vs. Massive/Gigantic MIMO}
The use of multiple antennas at both the transmitter and receiver, known as \gls{MIMO} technology, has gained significant popularity over the past decades due to its substantial performance-enhancing capabilities. MIMO technology offers numerous advantages over conventional \gls{SISO} systems, helping to address the challenges posed by wireless channel impairments and stringent resource constraints, such as power and bandwidth limitations. In addition to the time and frequency dimensions, which are the natural dimensions of digital communication, \gls{MIMO} technology leverages the spatial dimension by using multiple spatially distributed antennas.


Over the last few years, the use of very large antenna arrays (with a few hundred antennas) at the transmitter or receiver, named \textit{massive} \gls{MIMO}, has appeared as a promising candidate technology for increasing wireless throughput, reliability and energy efficiency \cite{LarssonMassiveMIMO2014,MarzettaMassiveMIMO2015,Marzetta16_massiveMIMO}. Massive \gls{MIMO} offers several advantages over the traditional configurations, such as \textit{channel hardening}. This effect appears as the number of antennas grows large and diminishes the impact of small-scale fading, so the channel behaves more like a deterministic channel as the number of antennas increases. Regardless of the channel hardening, the receiver still needs to obtain instantaneous CSI and perform coherent detection of the transmitted data based on it. Estimating the \gls{CSI} is generally carried out in a training phase, the complexity of which grows with the number of transmit antennas, thus being especially problematic in the downlink channel. To overcome this limitation, a noncoherent approach could be employed, where neither the transmitter nor the receiver has access to instantaneous \gls{CSI}. These schemes enable communication and exploit channel hardening through, at most, statistical \gls{CSI}, i.e., knowledge of the channel and noise distributions.






The upcoming sixth-generation (6G) communication system is set to harness the evolved \gls{MIMO} technology, referred to as \textit{gigantic} \gls{MIMO} \cite{LiGiganticMIMO2023,bjornson2024enabling6gperformanceupper}. This advancement is expected to support antenna arrays with over a thousand elements in low and mid frequency bands, and extend to thousands in high frequency bands such as \gls{THz}. With such an extensive number of antennas, gigantic MIMO has the potential to enable innovative use cases and applications that demand exceptionally high performance. For these new scenarios, noncoherent communication schemes appear as an effective solution to avoid the overhead imposed by channel estimation.


\subsubsection{Block-Fading vs. Continuously Varying Fading}

Wireless channels can be classified depending on how the fading coefficients vary. When the fading process is continuous with no abrupt temporal variations, the channel is said to be continuously varying. This model accurately represents mobile communications in indoor nomadic environments, rural broadband or backhaul links and in communication systems mounted on vehicles with continuous data transmission. In all these scenarios, the channel conditions remain relatively stable between consecutive channel uses, although significant changes may accumulate over extended timescales.



In contrast to continuously varying channels, in block-fading regimes the channel undergoes abrupt variations between blocks. This scenario appears when consecutive packets take independent paths to the receiver, and the channel in each path remains stable. In such cases, the channel can be approximated to be constant for a coherence interval of $T$ symbol periods and then take on an independent realization. This channel model arises in vehicle-to-vehicle and airborne communication networks, which are expected to be exceedingly agile to manage high mobility and rapid variations in channel conditions and demand. 


Block-fading channels also emerge in \gls{IoT} systems, whose purpose is the interconnection and management of intelligent devices with almost no human intervention. In these environments, communications are typically bursty, which leads to independent channels being observed over consecutive bursts, with the coherence time being a deterministic value corresponding to the burst duration. The evolved form of IoT for industrial environments, named \gls{IIoT}, has garnered significant attention as a key enabler for numerous potential vertical applications that will impact both business and society as a whole. Examples of these new applications are smart-grid energy management, smart cities, interconnected medical systems and autonomous drones \cite{mumtaz_iiot}. In these scenarios, wireless networks are typically characterized by a dense network of low-power, low-complexity terminals, the majority of which remain inactive most of the time. The information transmitted through these networks primarily consists of control and telemetry data collected by machine sensors. Despite the low traffic volume, the transmission must be highly reliable and meet stringent latency constraints. 



\subsection{Noncoherent Communication Schemes}

Applicable to different channel models and number of antennas at the receiver side, different noncoherent transmission schemes have been proposed in the literature. As explained in Sec. \ref{sec:intro:approaches}, we will focus here on three main approaches: 
Grassmannian signaling, differential-detection-based schemes, and energy-based schemes. 



\subsubsection{Grassmannian Signaling Schemes}

To achieve spectrally efficient communication over noncoherent block-fading channels, the transmitted signals must be tailored to the operating \gls{SNR}. At high \gls{SNR}s, it was demonstrated in~\cite{ZhengTse2002Grassman} that the optimal signaling technique consists of transmitting matrices representing points on the Grassmann manifold, which is a special case of the general noncoherent block-signaling structure proposed in~\cite{Hochwald2000unitaryspacetime}. For moderate-to-high \gls{SNR}s, the optimal signal structure remains unknown. However, in this regime, Grassmannian signaling is close to optimal, as shown in \cite{Hassibi2002_received_signal_density}.

The basic idea of Grassmannian signaling schemes is to encode the messages to be transmitted into different subspaces. When the coherence time of the channel is greater than the number of transmit antennas, the transmitted subspaces (represented by a semi-unitary or Stiefel matrix composed of orthogonal unit-norm columns) are invariant to the \gls{MIMO} channel, allowing the receiver to decode the received signal without the need of knowing the channel. Thus, this transmission technique is suited for MIMO systems of moderate size.



Since most communication systems usually operate in moderate-to-high or high \gls{SNR} regimes, Grassmannian signaling offers valuable performance advantages over training-based schemes for channels with short coherence time intervals like the ones that arise in vehicular communications or \gls{IoT} scenarios.

\subsubsection{Energy-Based Schemes}

With the goal of enabling \gls{HRLLC} for \gls{IIoT} systems, several energy-based schemes have been proposed for massive \gls{SIMO} systems \cite{PopovskiTSP16,aniol_ED,mallik_TCOM,hanConstellationDesignEnergyBased2022,xie_ED,Vucetic19,Manolakos2016noncoh_Energybased_massiveSIMO}. These schemes use energy-based modulations, where the information is only conveyed on the amplitude of the transmitted signal. For instance, \gls{PAM} is used in \cite{aniol_ED}, a multilevel \gls{ASK} constellation is used in \cite{mallik_TCOM}, or even tailored constellations for correlated channels are used in \cite{hanConstellationDesignEnergyBased2022,xie_ED}. All these works are based on the energy-detector scheme \cite{kay_ED}, which uses the squared norm of the received signal as its statistic. Note that this statistic is sufficient only when considering large antenna arrays at the receiver, due to the channel hardening effect. The energy detector is optimal under uncorrelated Rayleigh fading and allows for low-complexity implementations. However, its use under non-isotropic fading is suboptimal and its performance is known to be severely degraded by channel correlation \cite{sharma_ED,aniol_ED}.


\subsubsection{Differential-Detection-Based Schemes}
Differential MIMO signaling, akin to \gls{DPSK}, is particularly advantageous in scenarios where the channel fading process is continuous with no abrupt temporal variations and when channel estimation and tracking are deemed complex and challenging. The core concept of schemes based on differential detection involves encoding transmitted messages within the consecutive transitions between signaling blocks. These transitions can be reliably detected at the receiver, assuming that channel variations are almost negligible between blocks, but may accumulate substantial change over a longer period \cite{Baeza2018noncoh_SIMO_MAC_DPSK_BICM,victorTVT19,Shokrollahi_TIT01}. 



The differential-detection-based schemes proposed in \cite{Baeza2018noncoh_SIMO_MAC_DPSK_BICM,baeza2019noncoherent,victorTVT19} use a standard single-antenna DPSK modulation since they are intended for SIMO systems. In DPSK modulation, information is only conveyed in the phase of the transmitted signal. Assuming that the channel remains constant for two consecutive symbols, the phase difference between them can be detected at each receive antenna. 

The works in \cite{HochwaldDiff00,Hughes2000differentialSTM,Shokrollahi_TIT01} propose several multi-antenna differential modulation techniques for MIMO channels. Multi-antenna differential modulation is formally similar to standard single-antenna DPSK. In standard DPSK, the transmitted symbol has unit-modulus and is the product of the previously transmitted symbol and the current data symbol. As a generalization, $M$-antenna differential unitary space-time modulation differentially encodes $M \times M$ unitary matrix-valued signals. Essentially, these matrices are the product of the previously transmitted matrix and a unitary data matrix selected from a constellation $\mathcal{G}$, which forms a group under matrix multiplication. Consequently, all transmitted matrices also belong to $\mathcal{G}$. To enhance performance, differential MIMO schemes can be used in conjunction with efficient multiple-symbol detection schemes that incorporate trellis, sphere, and decision feedback decoding.



\begin{figure}[t!]
    \centering
    \includegraphics[width=\linewidth]{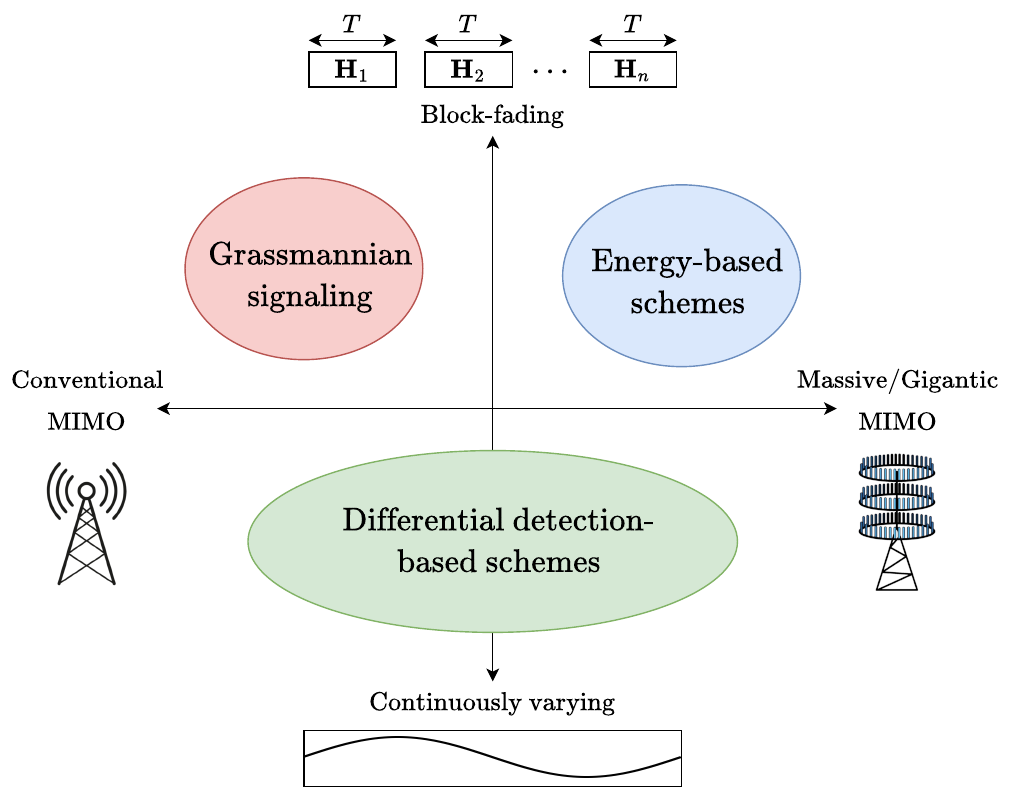}
    \caption{Best-suited noncoherent techniques for different scenarios. 
    }
    \label{fig:techniques_vs_scenarios}
\end{figure}

An overview of the different noncoherent communication design approaches that are addressed in this survey and their potential application scenarios is shown in Fig. \ref{fig:techniques_vs_scenarios}. We next review the theoretical foundation and existing designs of each approach.

\section{Grassmannian Signaling} \label{sec:Grassmannian}
In this section, we review the theoretical foundation and existing designs of Grassmannian constellations in both the single-user and multi-user scenarios. 

\subsection{Theoretical Foundation \note{ -Hoang}}
            
        



    Grassmannian signaling is inspired by the information-theoretic (near-)optimal input distribution for a Rayleigh block fading channel. Let us consider a \gls{MIMO} Rayleigh block fading channel with $\Nt$ transmit antennas and $\Nr$ receive antennas. The channel matrix $\Hm \in \CC^{\Nt \times \Nr}$ has \gls{iid} entries following $\Cc\Nc(0,1)$.\footnote{We denote by $\Cc\Nc(\muv, \Sigmam)$ the proper complex normal distribution with mean $\muv$ and covariance matrix $\Sigmam$.} It remains unchanged during a coherence block of $T \ge 2$ symbols and varies independently between blocks. Furthermore,~$\Hm$ is unknown to both the transmitter and the receiver. Let $\Xm \in \CC^{T \times \Nt}$ be the transmitted signal in a coherence block. The received signal is given by 
\begin{align} \label{eq:MIMO_block_fading}
    \Ym = \Xm \Hm + \sqrt{\frac{M}{TP}} \Zm,
\end{align}
where $\Zm$ is the \gls{AWGN} with independent entries following $\Cc\Nc(0,1)$. We assume that $\E[\|\Xm\|^2] \le M$, and with the normalization considered in~\eqref{eq:MIMO_block_fading}, $P$ is the \gls{SNR}. In the following, we assume 
that $M \le N$. Note that in the high-SNR regime, using more transmit antennas than receiver antennas does not improve the channel capacity~\cite{ZhengTse2002Grassman}.
    
    \subsubsection{Unitary Space-Time Modulation}
While Gaussian signaling is optimal in the coherent setting where $\Hm$ is known, it is suboptimal in the noncoherent setting (its achievable rate was evaluated in~\cite{Rusek2012mutual_information_Gaussian_signals,Alfano2014closed_form_output_stats}). 
In the high \gls{SNR} regime, a near-optimal input distribution is given by \gls{USTM}~\cite{Hochwald2000unitaryspacetime,Hassibi2002_received_signal_density}, in which $\Xm$ is isotropically distributed (i.e., its probability density is unchanged when pre-multiplied by a unitary matrix) and truncated unitary (i.e., $\Xm^\H \Xm = \Id_\Nt$). 
For $T \ge M + N$, \gls{USTM} was shown to achieve the channel capacity up to a vanishing gap as the \gls{SNR} $P$ goes to infinity~\cite{ZhengTse2002Grassman}. For $2M \le T \le M+N$, the high-\gls{SNR} capacity-achieving input distribution requires augmenting \gls{USTM} with power control over the antennas 
where the power terms allocated to the antennas have the same joint distribution as the eigenvalues of a positive-definite $\Nt \times \Nt$ Beta-distributed random matrix~\cite{Yang2013CapacityLargeMIMO}.\footnote{In fact, for every \gls{SNR} value, the capacity-achieving input has the form of a product of a $T \times M$ isotropically-distributed truncated-unitary matrix and an $M\times M$ real, nonnegative, diagonal matrix~\cite{Marzetta1999capacity}.} However, the rate achievable with \gls{USTM} is only at a constant gap below the capacity~\cite[Coro.~5]{Yang2013CapacityLargeMIMO}.
Finally, for $1 < T < 2M$, \gls{USTM} achieves the optimal \gls{DoF}, i.e., pre-log factor  of the capacity
~\cite[Sec.~IV-D]{ZhengTse2002Grassman}.

In the finite blocklength regime, where a channel codeword spans a finite number of coherence blocks, a more relevant metric than the ergodic capacity is the achievable rate for a fixed error probability~\cite{Polyanskiy10,Durisi16toward}. 
The optimal signaling is not known, but 
\gls{USTM} has been used to derive best-known achievability bounds~\cite{Durisi16_FBL,Lancho2020SISOblockfading,Ostman2019pilotVSnoncoherent,Qi2025noncoherent,Lancho20saddlepoint}. These bounds show that \gls{USTM} outperforms the pilot-based scheme~\cite{Ostman2019pilotVSnoncoherent} and the Alamouti space-time modulation combined with frequency-switched transmit diversity~\cite{Durisi16_FBL}. 

\subsubsection{Grassmannian Constellations}
With \gls{USTM}, 
information is embedded in the column space of the matrix $\Xm$. In other words, information is carried in the position of the column space of $\Xm$ in the Grassmann manifold $\Gras$. The intuition behind the near-optimality of \gls{USTM} is that the channel matrix~$\rvMat{H}$ only scales and rotates the basis of the transmitted signal matrix~$\Xm$ without changing its column space. Indeed, the column spaces of $\Xm$ and the noise-free observation~$\Xm\rvMat{H}$ are identical; they therefore represent the same element of $\Gras$. At high \gls{SNR}, the additive noise has a low impact on the subspace of the output signal, and the column space of~$\Xm$ can be accurately recovered from the column space of the noisy output $\Ym$. 
From this observation, a constellation for the channel~\eqref{eq:MIMO_block_fading} 
can be designed as a set of 
points on the Grassmann manifold $\Gras$, i.e., $\Xm$ is drawn from
\begin{align}
\Xc \defeq \big\{\Xm_1,\dots,\Xm_{L} \in \CC^{T\times M} \colon \Xm_i^\H\Xm_i = \Id_M, i\in [L] \big\}
\label{eq:grass_constellation}
\end{align}
which contains representatives of $L$ subspaces in $\CC^T$.
We refer to constellations such as $\Xc$ as \emph{Grassmannian constellations}.

The design of a Grassmannian constellation is essentially a point packing problem in the Grassmann manifold. We present a brief tutorial on packing in Riemannian manifolds, particularly the Grassmann manifold, in Appendix~\ref{app:manifold}, and show minimum chordal distances for different optimized packings in  Appendix~\ref{app:best_packings}. There are various geometric distance measures between points on the Grassmann manifold, such as the chordal distance, geodesic distance, and Fubini-Study distance. However, for constellation design, it is important to choose a packing criterion related to the communication performance. This performance is typically captured by the average detection error probability of the optimal \gls{ML} detector. 
This probability is upper and lower bounded, up to constant multiplicative factors, by the maximum of the \gls{PEP} over all symbol pairs of constellation symbols~\cite[Sec.~III]{NgoAsilomar2018multipleAccess}. 
Here, the \gls{PEP} $\P[\Xm_i \to \Xm_j]$ is the probability of detecting $\Xm_j$ while $\Xm_i$ was transmitted. As a consequence, design criteria can be derived from an analysis of the \gls{PEP}, as we present next.

\begin{itemize}[leftmargin=*]
    \item \emph{Maximizing the minimum pairwise chordal distance:}  \cite[Th.~5]{Hochwald2000unitaryspacetime} showed that $\P[\Xm_i \to \Xm_j]$ decreases as any singular value of the matrix $\Xm_i^\H\Xm_j$ decreases. A resulting design criterion is to minimize the largest among such singular values~\cite[Sec.~VI-B]{Hochwald2000unitaryspacetime}. Building up on~\cite[Th.~5]{Hochwald2000unitaryspacetime},\cite{Agrawal2001MIMOconstellations} proposes to use the sum of the squared singular values, which is equal to $\trace(\Xm_i^\H \Xm_j \Xm_i^\H \Xm_j)$, as the quantity that governs the \gls{PEP}. This leads to the design criterion of {maximizing the minimum chordal distance} $d(\Xm_i,\Xm_j) \defeq \sqrt{M-\trace(\Xm_i^\H \Xm_j \Xm_i^\H \Xm_j)}$ between all pairs of constellation symbols $\Xm_i$ and $\Xm_j$.
    Leveraging a Taylor expansion of the \gls{PEP} around $P = 0$, \cite{CuevasThesis2024} shows that this criterion is also optimal in the low-\gls{SNR} regime. This connection with the chordal distance allows benchmarking the packing efficiency with sphere-packing bounds derived in, e.g.,~\cite{Rankin_1955,Conway1996packing,Barg2002BoundsOP,Henkel2005spherePackingBounds,Dai2008quantizationBounds}.


    \item \emph{Minimizing the asymptotic \gls{PEP}:} \cite[Prop.~4]{Brehler2001asymptotic} showed that, in the high-\gls{SNR} regime, \gls{PEP} approaches asymptotically to a term proportional to $\det^{-N} (\Id_M - \Xm_i^\H \Xm_j \Xm_i^\H \Xm_j)$. Based on this result, the authors of~\cite{McCloudIT2002signalDesignAndConvCode} invoked the union bound to propose a design criterion consisting in minimizing the sum of $\det^{-N} (\Id_M - \Xm_i^\H \Xm_j \Xm_i^\H \Xm_j)$ over all symbol pairs $(\Xm_i,\Xm_j)$. Notably, this criterion depends on the number of receive antennas $N$.
    Removing this dependency, the authors of \cite{Alvarez22coherence} proposed a criterion consisting in maximizing the minimum value of $\det(\Id_M - \Xm_i^\H \Xm_j \Xm_i^\H \Xm_j)$ over all symbol pairs. 

    \item \emph{Maximizing the \gls{KL} divergence:} The \gls{PEP} $\P[\Xm_i \to \Xm_j]$ can be seen as the type-$1$ error probability of the likelihood ratio test of the per-antenna received signal to be distributed as $\Cc\Nc(\mathbf{0}, \Id_T + \Xm_i \Xm_i^\H)$ or $\Cc\Nc(\mathbf{0}, \Id_T + \Xm_j \Xm_j^\H)$. An application of the Chernoff-Stein lemma shows that the optimal asymptotic ($N \to \infty$) error exponent of $\P[\Xm_i \to \Xm_j]$ is given by the \gls{KL} divergence between $\Cc\Nc(\mathbf{0}, \Id_T + \Xm_i \Xm_i^\H)$ and $\Cc\Nc(\mathbf{0}, \Id_T + \Xm_j \Xm_j^\H)$.\footnote{This error exponent is not achieved by the \gls{ML} detector, but its relation to the \gls{ML} detection error probability has been pointed out in~\cite[Lem.~3]{BorranTIT2003nonCoherentKLdistance}.} Therefore, the constellation can be designed by maximizing the minimum value of the mentioned \gls{KL} divergence over all symbol pairs~\cite{BorranTIT2003nonCoherentKLdistance,Ngo2022joint_constellation_design}.  
\end{itemize}

A criterion not directly derived from the \gls{PEP} was proposed in~\cite{Gohary2009GrassmanianConstellations}. The authors analyzed the effect of the noise on the signal subspace and proposed to maximize the minimum pairwise chordal Frobenius norm $d_{\rm F}(\Xm_i,\Xm_j) \defeq \sqrt{2(M-\trace(\Xm_i^\H \Xm_j))}$.

\subsection{Single-User Constellation Design \note{- Diego}}



Existing Grassmannian constellation designs can be generically categorized into two groups: unstructured and structured. The first group uses numerical optimization tools to solve the packing problem on the Grassmannian, whereas the second imposes a particular structure on the constellation to facilitate low complexity mapping and/or demapping.

\subsubsection{Unstructured Constellations} 

 Unstructured constellations are obtained from a numerical solution of the packing problem on the Grassmann manifold according to a certain distance metric or cost function, such as the ones mentioned in the previous subsection. 
 Since the constellation points are unitary matrices without any particular structure, it is necessary to store the full constellation and the set of binary labels at both the transmitter and the receiver to perform the bit-to-symbol mapping/demapping as well as to implement the detector. Due to the lack of low-complexity detectors applicable to this type of Grassmannian constellations, it is necessary to employ the optimal \gls{ML} detector. 
 This detector computes the projections of the received signal on all $L= 2^{RTM}$ constellation points, so its computational cost grows exponentially with the transmission rate $R$ in bits/antenna/channel use. For these reasons, the use of unstructured constellations is limited in practice to relatively low transmission rates, e.g., $R<1.5$ bits/antenna/channel use. As an example, for a system with $M=2$ transmit antennas and coherence time $T=4$ the cardinality of a constellation with rate $R=1.5$ is $L = 4096$ points.


 Most classic design methods for unstructured constellations aim at maximizing the minimum distance between constellation points, using different suitable metrics. For instance, the numerical methods in \cite{Agrawal2001MIMOconstellations,Gohary2009GrassmanianConstellations,Peng2017grassmannian} employ the chordal distance while \cite{BekoTSP2007noncoherentColoredNoise} uses the spectral distance. The numerical method in \cite{Dhillon2008constructing}, which enforces in each iteration both structural and spectral properties of the Gram matrix\footnote{Considering a set of points on the Grassmann manifold represented as in \eqref{eq:grass_constellation}, the Gram matrix is defined in \cite{Dhillon2008constructing} as the $L \times L$ block matrix composed of $M \times M$ blocks, where each block is given by $\Gm_{ij} = \Xm_i^\H\Xm_j$, $i = 1,\dots,L$, $j = 1, \dots, L$.} formed by the inner products between constellation points, provides a more flexible optimization framework and may use the chordal distance, the spectral distance, or the Fubini-Study distance (see Appendix \ref{app:manifold}). However, this method has several drawbacks, such as its slow convergence in certain cases, due mainly to the increase of the size of the Gram matrices involved in the optimization with the number of antennas at the transmitter $M$ and/or the constellation size $L$. In addition, it may not deliver good packings when the ambient dimension of the Grassmann manifold is large.


 Some other classic constellation design algorithms aim at minimizing the asymptotic \gls{PEP}, invoking the union bound such as \cite{McCloudIT2002signalDesignAndConvCode,Wu2008USTM_based_on_Chernoff_bound} or just focusing on the dominant term over all symbol pairs (the so-called \textit{diversity product}) \cite{Han06}. The problem of maximizing the diversity product is equivalent to minimizing the \textit{coherence} of a configuration of subspaces\footnote{Notice that if we consider single-antenna designs, this criterion is also equivalent to maximizing the minimum chordal distance.} as proposed in \cite{Alvarez22coherence,Tahir2019constructing_Grassmannian_frames}. In \cite{BorranTIT2003nonCoherentKLdistance}, the optimization criterion is the maximization of the KL divergence, which is related to the error exponent of the asymptotic \gls{PEP}.

 In recent years, Grassmannian signaling has regained significant interest, and some new constellation designs have appeared in the literature. In~\cite{Cuevas21WSA}, a fast algorithm for maximizing the minimum chordal distance while attaining top-tier error rates is proposed. This optimization framework has been later extended to consider the maximization of the diversity product in \cite{Alvarez22coherence} and the minimization of the asymptotic \gls{PEP} union bound in~\cite{Cuevas23unionBound}. Another recent approach that considers the minimization of the coherence between subspaces in the Grassmannian of lines, i.e. when considering single-antenna transmitters, has been proposed in \cite{Park22Allerton}. Here, a new optimization algorithm is proposed with a CPU/GPU parallelized implementation that offers superior performance over existing methods when the size of the matrices involved is large. 

 Some other recent approaches have considered data-driven deep learning techniques to solve the constellation design problem. Among them, the methods based on autoencoders for Grassmannian constellation design in \cite{Fu2021grassmannian,Xiaotian2021deep_learning} stand out. Another work that has used deep learning techniques is \cite{Baba24}, in which a reduced-complexity deep neural network-based detector for Grassmannian constellations is proposed.

 Very recently, Grassmannian constellations have been proposed to be used as data-carrying reference signals~
 \cite{Endo24,kato2025maximizingspectrumefficiencydatacarrying}, which allow for simultaneous data transmission and channel estimation to achieve boosted spectral efficiency. In these papers, symbols on the Grassmann manifold are exploited to carry additional data and to assist in channel estimation for the coherent transmission phase. In \cite{Endo24}, a set of unitary matrices $\big\{\Um_1, \ \dots, \ \Um_L \big\}$ is numerically optimized to minimize the channel estimation error of an existing Grassmannian constellation. Notice that, by definition, points on the Grassmann manifold are invariant to right rotations, so the error rate performance of the new constellation $\Xc = \big\{\Xm_1\Um_1, \ \dots, \ \Xm_{L}\Um_L \big\}$ remains the same as that of the original constellation $\big\{\Xm_1, \ \dots , \ \Xm_{L} \big\}$. The work in \cite{Endo24} has been later extended in \cite{kato2025maximizingspectrumefficiencydatacarrying} to consider a numerical optimization-based Grassmannian constellation design that accounts for both data transmission and channel estimation. Here, an upper bound on the normalized mean squared error of the estimated channel matrices and a lower bound on the noncoherent average mutual information are optimized simultaneously by using a Bayesian optimization technique.
 

For clarity and ease of comparison, Table \ref{tab:grassmann_designs_unstructured} provides a summary of the previously discussed papers about unstructured Grassmannian constellations, emphasizing their main contributions.

\begin{table*}[t!]
\renewcommand{\arraystretch}{1.1}
\caption{Literature overview on single-user unstructured Grassmannian constellation design}
\centering
\begin{tabular}{ |m{3.5cm}|m{.7cm}|m{1cm}|m{3.5cm}|m{7.15cm}| } 
\hline
\textbf{Design} & \textbf{Year} & \textbf{Scenario} & \textbf{Criterion} &\textbf{Algorithm} \\
\hline \hline
Agrawal \textit{et al.} \cite{Agrawal2001MIMOconstellations} & 2001 & MIMO & Chordal distance & Use of surrogate functionals to solve a sequence of relaxed nonlinear programs\\
\hline
McCloud \textit{et al.} \cite{McCloudIT2002signalDesignAndConvCode} & 2002 & MIMO & Asymptotic union bound (AUB) on probability of error & 
Gradient-based optimization + greedy approach (optimizes a new constellation point each time)
\\
\hline
Borran \textit{et al.} \cite{BorranTIT2003nonCoherentKLdistance} & 2003 & MIMO & KL divergence & Optimization of smaller subsets using sequential quadratic programming methods\\
\hline
Han and Rosenthal \cite{Han06} & 2006 & MIMO & Diversity product / Diversity sum & Simulated annealing algorithm\\
\hline
Beko \textit{et al.} \cite{BekoTSP2007noncoherentColoredNoise} & 2007 & MIMO & Spectral distance & Convex semidefinite programming relaxation of design problem + refinement through geodesic descent algorithm\\
\hline
Dhillon \textit{et al.} \cite{Dhillon2008constructing} (Alternating projection) & 2008 & MIMO & Chordal distance, spectral distance or Fubini-Study distance & Enforcement of structural and spectral properties of Gram matrix formed by inner products between constellation points\\
\hline
Wu \textit{et al.} \cite{Wu2008USTM_based_on_Chernoff_bound} & 2008 & MIMO & \gls{PEP} Chernoff bound & Gradient descent search on a family of surrogate functions\\
\hline
Gohary and Davidson \cite{Gohary2009GrassmanianConstellations} & 2009 & MIMO & Chordal distance & Gradient-based algorithm on a higher-dimensional Grassmann manifold\\
\hline
Peng \textit{et al.} \cite{Peng2017grassmannian} & 2017 & MIMO & Chordal distance & Grid search of antipodal orthogonal constellation\\
\hline
Tahir \textit{et al.} \cite{Tahir2019constructing_Grassmannian_frames} & 2019 & SIMO & Coherence / Cross-correlation & Iterative collision-based packing of equal-radius hyperspheres on the surface of a unit-norm hypersphere\\
\hline
Fu and Le Ruyet \cite{Fu2021grassmannian,Xiaotian2021deep_learning} & 2021 & MIMO & Input/output cross-entropy & Use of autoencoders for Grassmannian constellation design (deep learning approach)\\
\hline
Cuevas \textit{et al.} \cite{Cuevas21WSA} & 2021 & MIMO & Chordal distance & Gradient ascent optimization on the Grassmann manifold\\ 
\hline
Álvarez-Vizoso \textit{et al.} \cite{Alvarez22coherence} & 2022 & MIMO & Coherence / Diversity product & Gradient ascent optimization on the Grassmann manifold\\
\hline
Park \textit{et al.} \cite{Park22Allerton} & 2022 & SIMO & Coherence / Cross-correlation & Alternating method of computing successive smooth approximations to the maximum function coupled with an unconstrained minimization procedure\\
\hline
Cuevas \textit{et al.} \cite{Cuevas23unionBound} & 2023 & MIMO & \gls{PEP} union bound & Gradient ascent optimization on a product of Grassmannian manifolds\\
\hline
Cuevas \textit{et al.} \cite{Cuevas24harware} & 2024 & MIMO & Coherence / Diversity product & Gradient ascent optimization on the Grassmann manifold considering carrier frequency offset and I/Q imbalance perturbed constellation points\\
\hline
Endo \textit{et al.} \cite{Endo24} & 2024 & MIMO & Channel estimation accuracy & Manifold optimization techniques (could also be used in combination with structured constellations)\\
\hline
Kato \textit{et al.} \cite{kato2025maximizingspectrumefficiencydatacarrying} & 2025 & MIMO & Channel estimation accuracy \& Spectral efficiency & Bayesian optimization technique\\
\hline 
\end{tabular}
\label{tab:grassmann_designs_unstructured}
\end{table*}
        
\subsubsection{Structured Constellations} 

\begin{figure}[!t] 
	\centering
	\includegraphics[width=.48\textwidth,trim=1cm 3.7cm 1cm 1cm,clip=true]{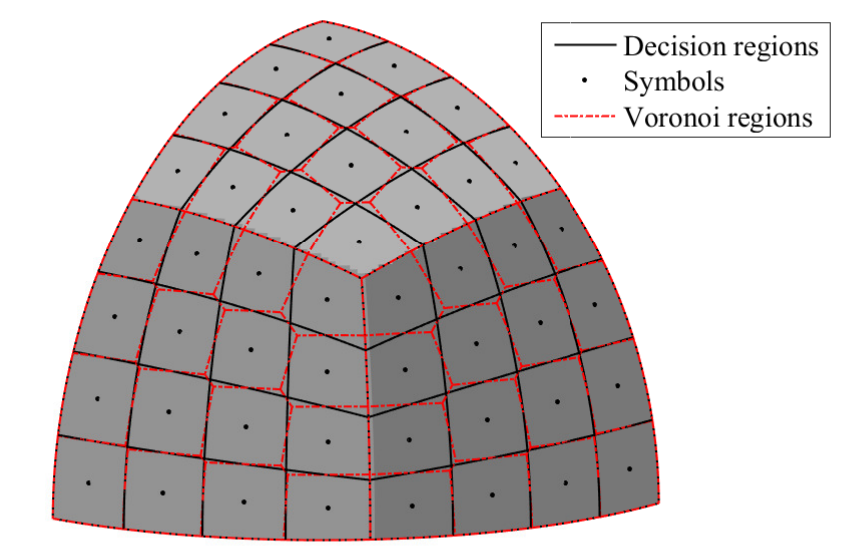}
	\caption{Illustration of the symbols and decision regions of the greedy decoder for a section (around the cell boundaries) of the Cube-Split constellation~\cite{Ngo_cubesplit_journal} on $\GG(1,\RR^3)$ for $M = 1$ antenna and spectral efficiency $\eta = R/M = 5$ bits/s/Hz. Thanks to the regular placement of the symbols, these decision regions well match the Voronoi regions, which are the optimal decision regions.
	}
	\label{fig:decisionRegions}
\end{figure}

To simplify the symbol detection problem, especially for moderate to high transmission rates, it is necessary to introduce some structure on the constellation.  Structured constellations eliminate or reduce the need to store all constellation points in the transmitter and receiver (allowing, for example, to apply a bit-to-symbol mapper on the fly) and/or facilitate low complexity detection. A well-designed structure can enable the low-complexity detector to approach the performance of the ML detector; see an illustration in Fig.~\ref{fig:decisionRegions}. As a general rule, structured constellations are necessary for high transmission rates $R>2$ bits/antenna/channel use. Nevertheless, it is important to bear in mind that structured constellations provide packings with worse separation between points than numerically optimized unstructured ones, which necessarily translates into poorer performance in terms of \gls{SER} or \gls{BER}.

 Existing structured designs impose some kind of structure on the constellation points through algebraic constructions such as the Fourier-based constellation in \cite{Hochwald2000systematicDesignUSTM}, which constructs the desired constellation by successive rotations of an initial signal in a high-dimensional complex space, entailing the imposition of a circulant correlation structure; the designs based on the use of the Cayley transform in \cite{JingTSP2003Cayley}; the designs based on group representations in \cite{Pitaval12}, which construct the constellation from an initial point by using finite groups with unitary representations; and the designs based on the augmentation of a small close-to-optimal constellation by using geodesics proposed in \cite{AttiahISIT2016systematicDesign}.

 Another popular way of generating structured Grassmannian constellation designs that has been used in the literature is leveraging coherent space-time codes and constellations. Among these, we can mention the generalization of \gls{PSK} constellations in \cite{TarokhIT2002existence}; the use of training codes in \cite{DayalIT2004leveraging}; the Exp-Map parameterized mapping of unitary matrices in \cite{Kammoun2007noncoherentCodes}, which uses a non-linear map, called the exponential map, applied to space-time codes for coherent systems to construct the constellation; and the designs based on the QR decomposition of coherent nonunitary space-time block-codes based on unique factorizations of pairs of \gls{PSK} constellations in \cite{Zhang2011full_diversity_blind_ST_block_codes}.

 Among the recently proposed structured Grassmannian constellation designs, we can mention the uniquely factorable constellations in \cite{LiDesign2019}, which ensure the unique identification of the transmitted symbols in the noise-free case in \gls{SIMO} systems. The space-time block codes based on \gls{QEC} in \cite{Cuvelier21quantum} use the algebraic structure of \gls{QEC} codes to map numerically-optimized Grassmannian packings in lower-dimensional spaces to codewords in a higher-dimensional space. The analog subspace codes proposed in \cite{Soleymani22} leverage polynomial evaluations over finite fields that map their elements to the unit circle in the complex plane in order to correct errors/erasures over the analog operator channel \cite{KoetterCoding2008}. 
 
 Another recent constellation design is the Cube-Split method for SIMO channels in \cite{Ngo_cubesplit_journal}, which uses structured partitions of the Grassmannian. The Cube-Split constellation is generated by partitioning the Grassmannian of lines into a collection of bent hypercubes or cells and defining a mapping onto each of these bent hypercubes such that the resulting symbols are approximately uniformly distributed on the Grassmannian. The Grassmannian is partitioned by assigning each cell to the Voronoi region associated with a canonical basis vector. Then, a grid in the Euclidean space of the same real dimensions as the considered Grassmann manifold, i.e. $2(T-1)$ real dimensions, is defined. For each real component, the grid is composed of points regularly spread on the interval $(0,1)$ in order to maximize the minimum distance within the set. Finally, the uniformly distributed points in the Euclidean space are mapped onto uniformly distributed points on each cell. This structure enables a simple greedy detector whose decision regions are close to the Voronoi regions of the symbols, which are the decision regions of the optimal ML detector, as illustrated in Fig.~\ref{fig:decisionRegions}. However, when $T>2$, Cube-Split ignores the statistical dependencies between the components of the constellation points and applies the same mapping derived for $T=2$ in a component-wise manner. Thus, the constellation points for $T > 2$ are not uniformly distributed on each cell.

\begin{figure}[t!]
     \centering
     \input{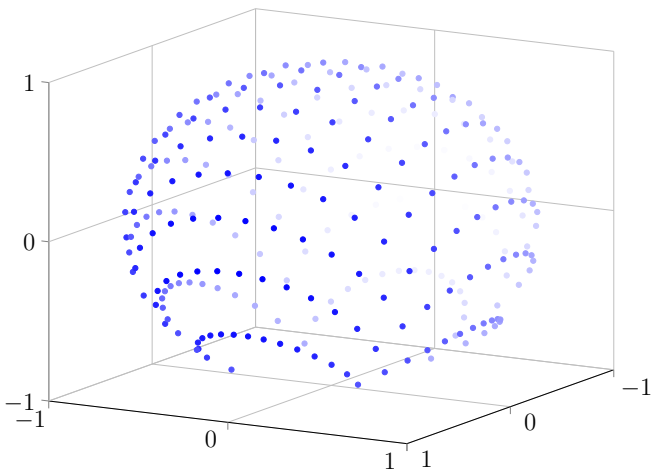}
     \caption{Representation of Grass-Lattice constellation for $T = 2$ symbol periods, $M = 1$ antenna and spectral efficiency $\eta = R/M = 4$ bits/s/Hz from \cite{Cuevas24constellationsOnTheSphere}, where the Hopf map \cite{hopfmap} is used to represent the points on the Grassmann manifold.}
     \label{fig:grasslattice_mapping}
 \end{figure}

 To overcome this problem, a new mapping, called Grass-Lattice, was proposed in \cite{Cuevas2022measure,Cuevas24constellationsOnTheSphere}, which establishes a measure-preserving transformation between the unit hypercube and the Grassmannian for any $T \geq 2$. The measure-preserving nature of the Grass-Lattice mapping ensures that any set of points uniformly distributed in the input space (the hypercube) is mapped onto a corresponding set of points uniformly distributed in the output space (the Grassmann manifold) \cite{Beltran2025MeasurePreserving}. The proposed constellation design is derived from the composition of three sequential mappings that together form a measure-preserving transformation. The first mapping converts points uniformly distributed in the unit hypercube into normally distributed points. The second mapping then maps these normally distributed points to points uniformly distributed within the unit ball (subset of vectors with modulus less than one). Finally, the third mapping transforms these uniformly distributed points in the unit ball into points uniformly distributed in the Grassmannian of lines. An example of a constellation generated using the Grass-Lattice mapping for the case $T = 2$ and $M = 1$ is shown in Fig. \ref{fig:grasslattice_mapping}.

 While alternative methods, such as random Gaussian matrices~\cite{ShiRandomGaussianMatrices2015}, can be used to obtain uniformly distributed points in the Grassmannian, these constellations often lack structure and require an exhaustive detector. The structure of constellation points generated through Grass-Lattice and Cube-Split mapping facilitates on-the-fly symbol generation, low-complexity decoding, and simple bit-to-symbol Gray-like coding. Grass-Lattice mapping was later extended to the case of MIMO systems with $M = 2$ transmit antennas in \cite{Cuevas2024structured}.

 For clarity and ease of comparison, Table \ref{tab:grassmann_designs_structured} presents a summary of the previously discussed works on structured Grassmannian constellations, highlighting their main contributions.

\begin{table*}[t!]
\renewcommand{\arraystretch}{1.1}
\caption{Literature overview on single-user structured Grassmannian constellation design}
\centering
\begin{tabular}{ |m{4.5cm}|m{.7cm}|m{1cm}|m{10cm}| } 
\hline
\textbf{Design} & \textbf{Year} & \textbf{Scenario} & \textbf{Main idea} \\
\hline \hline
Hochwald \textit{et al.} \cite{Hochwald2000systematicDesignUSTM} (Fourier) & 2000 & MIMO & Successive rotations of initial signal in a high-dimensional complex space, entailing the imposition of a circulant correlation structure\\
\hline
Tarokh and Kim \cite{TarokhIT2002existence} & 2002 & MIMO & Constellation design based on generalization of \gls{PSK} constellations and orthogonal designs\\
\hline
Jing and Hassibi \cite{JingTSP2003Cayley} & 2003 & MIMO & Cayley transform-based constellation design\\
\hline
Zhao \textit{et al.} \cite{ZhaoTIT2004orthogonalDesign} & 2004 & MIMO & Orthogonal algebraic constructions combined with trellis coding\\
\hline
Dayal \textit{et al.} \cite{DayalIT2004leveraging} & 2004 & MIMO & Use of training codes with structure inspired by the pilot-based scheme\\
\hline
Kammoun \textit{et al.} \cite{Kammoun2007noncoherentCodes} (Exp-Map) & 2007 & MIMO & Application of non-linear exponential map to space-time codes for coherent systems\\ 
\hline
Ashikhmin and Calderbank \cite{Ashikhmin10} & 2010 & MIMO & Multidimensional generalizations of binary Reed-Muller codes\\
\hline
Zhang \textit{et al.} \cite{Zhang2011full_diversity_blind_ST_block_codes} & 2011 & MIMO & QR decomposition of nonunitary space-time block-codes based on unique factorizations of pairs of \gls{PSK} constellations\\
\hline
Pitaval and Tirkonnen \cite{Pitaval12} & 2012 & MIMO & Constellation construction from an initial point by using finite groups with unitary representations\\
\hline
Attiah \textit{et al.} \cite{AttiahISIT2016systematicDesign} & 2016 & MIMO & Augmentation of a small close-to-optimal constellation by using geodesics\\
\hline
Li \textit{et al.} \cite{LiDesign2019} & 2019 & SIMO & Unitary uniquely factorable constellations (UUFC)\\
\hline
Ngo \textit{et al.} \cite{Ngo_cubesplit_journal} (Cube-Split) & 2020 & SIMO & Structured partition of the Grassmannian of lines into a collection of bent hypercubes or cells and mapping onto each of them such that resulting symbols are approximately uniformly distributed on the Grassmannian\\
\hline
Cuvelier \textit{et al.} \cite{Cuvelier21quantum} & 2021 & MIMO & Use of \gls{QEC} codes to map numerically-optimized Grassmannian packings in lower-dimensional spaces to codewords in a higher-dimensional space\\
\hline
Soleymani and Mahdavifar \cite{Soleymani22} & 2022 & MIMO & Leverage of polynomial evaluations over the analog operator channel\\
\hline
Cuevas \textit{et al.} \cite{Cuevas24constellationsOnTheSphere} (Grass-Lattice) & 2024 & SIMO & Measure preserving mapping between the unit hypercube and the Grassmann manifold for 1 transmit antenna\\
\hline
Cuevas \textit{et al.} \cite{Cuevas2024structured} (Grass-Lattice) & 2024 & MIMO & Measure preserving mapping between the unit hypercube and the Grassmann manifold for 2 transmit antennas\\
\hline
\end{tabular}
\label{tab:grassmann_designs_structured}
\end{table*}

\subsubsection{Bit-Labeling}
A critical practical consideration in noncoherent communications employing Grassmannian constellations is the design of an effective bit-to-symbol mapping. In this setting, each constellation point needs to be assigned a binary label with the goal of ensuring that points in close proximity (or equivalently, those with high pairwise error probability) are mapped to binary sequences with small Hamming distance. In coherent communications using standard constellations arranged on a regular lattice (e.g, 64-QAM), the optimal mapping that minimizes the average \gls{BER} is achieved via Gray mapping, which guarantees that adjacent symbols differ by just a single bit. However, the binary labeling problem becomes considerably more challenging for Grassmannian constellations, particularly for unstructured ones. While optimal Gray-like mappings have been established for certain structured constellations \cite{ZhaoTIT2004orthogonalDesign,Nghi2007USTM_iterativeDecoding} and quasi-Gray schemes have been proposed in \cite{Qin20quasiGray} for structured codes such as Reed-Muller Grassmannian constellations \cite{Ashikhmin10}, unstructured constellations typically feature a number of neighboring symbols that exceeds the available binary labels, thereby precluding Gray mapping. Under these conditions, determining the optimal labeling for a constellation $\mathcal{X}$ of cardinality $L$ would require the definition of a cost function (e.g., average \gls{BER}) and an exhaustive search over $L!$ possible mappings. As a result, suboptimal labeling schemes that decouple the constellation design problem from the binary labeling task are frequently employed \cite{Yabo05mapping,Colman2011quasi-gray,Baluja2013neighborhood}. More recently, a joint binary labeling and constellation design algorithm was proposed in \cite{Cuevas23unionBound} by simply including weights in the \gls{PEP} union bound cost function that are proportional to the Hamming distance between codewords. Consequently, the cost function assigns a greater weight to pairs of codewords with larger bit differences, which compels the optimization process to position these codewords further apart compared to those with a smaller Hamming distance.


\subsubsection{Performance Comparison}

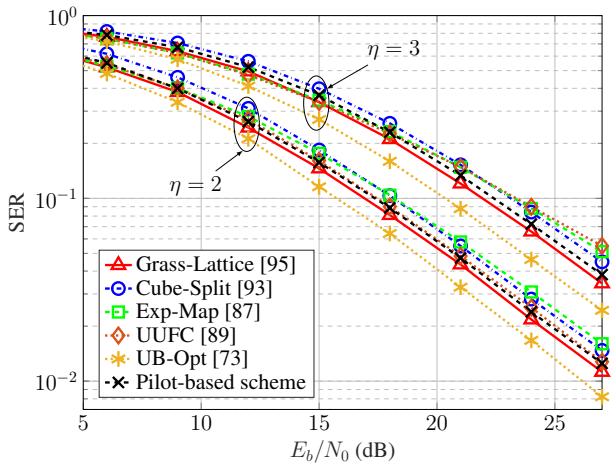
\begin{figure}[t!]
    \begin{center}
%
%
\definecolor{mycolor1}{rgb}{0.92941,0.69412,0.12549}%
\definecolor{mycolor2}{rgb}{0.85098,0.32549,0.09804}%
\resizebox{\columnwidth}{!}{
	
	\begin{tikzpicture}
		
		\begin{axis}[%
			width=4.58in,
			height=3.48in,
			at={(0.758in,0.564in)},
			scale only axis,
			font = \Large,
			xmin=5,
			xmax=27,
			xtick={5, 10, 15, 20, 25},
			xlabel style={font=\Large\color{white!15!black}},
			xlabel={$E_b / N_0$ (dB)},
			ymode=log,
			ymin=0.007,
			ymax=1,
			yminorticks=true,
			ylabel style={font=\Large\color{white!15!black}},
			ylabel={SER},
			axis background/.style={fill=white},
			xmajorgrids,
			ymajorgrids,
			minor grid style = {dashed},
			yminorgrids,
			legend style={at={(0.03,0.03)}, anchor=south west, legend cell align=left, align=left, draw=white!15!black}
			]
			\addplot [color=red, line width=1.5pt, mark size=5pt, mark=triangle, mark options={solid, red}]
			table[row sep=crcr]{%
				0	0.7706\\
				3	0.66995\\
				6	0.5209\\
				9	0.38175\\
				12	0.2438\\
				15	0.14586\\
				18	0.081085\\
				21	0.04349\\
				24	0.02175\\
				27	0.011242\\
			};
			\addlegendentry{Grass-Lattice \cite{Cuevas24constellationsOnTheSphere}}
			
			\addplot [color=blue, dashdotted, line width=1.5pt, mark size=4.0pt, mark=o, mark options={solid, blue}]
			table[row sep=crcr]{%
				0	0.84015\\
				3	0.7454\\
				6	0.61775\\
				9	0.45935\\
				12	0.3107\\
				15	0.183895\\
				18	0.103505\\
				21	0.054925\\
				24	0.02812\\
				27	0.01473\\
			};
			\addlegendentry{Cube-Split \cite{Ngo_cubesplit_journal}}
			
			\addplot [color=green, dashed, line width=1.5pt, mark size=3.5pt, mark=square, mark options={solid, green}]
			table[row sep=crcr]{%
				0	0.76895\\
				3	0.6707\\
				6	0.52935\\
				9	0.40195\\
				12	0.28045\\
				15	0.176375\\
				18	0.10416\\
				21	0.057735\\
				24	0.0307533333333333\\
				27	0.016022\\
			};
			\addlegendentry{Exp-Map \cite{Kammoun2007noncoherentCodes}}
			
			\addplot [color=mycolor2, dotted, line width=1.5pt, mark size=5pt, mark=diamond, mark options={solid, mycolor2}]
			table[row sep=crcr]{%
				3	0.66715\\
				6	0.5425\\
				9	0.39945\\
				12	0.2704\\
				15	0.16062\\
				18	0.090215\\
				21	0.048365\\
				24	0.02525\\
				27	0.012766\\
			};
			\addlegendentry{UUFC \cite{LiDesign2019}}
			
			\addplot [color=mycolor1, dotted, line width=1.5pt, mark size=5.0pt, mark=asterisk, mark options={solid, mycolor1}]
			table[row sep=crcr]{%
				0	0.75175\\
				3	0.64385\\
				6	0.4824\\
				9	0.3354\\
				12	0.2119\\
				15	0.115725\\
				18	0.064245\\
				21	0.03266\\
				24	0.0167166666666667\\
				27	0.008216\\
			};
			\addlegendentry{UB-Opt \cite{Cuevas23unionBound}}
			
			\addplot [color=black, dashed, line width=1.5pt, mark size=5.0pt, mark=x, mark options={solid, black}]
			table[row sep=crcr]{%
				0	0.7907\\
				3	0.67915\\
				6	0.55135\\
				9	0.39945\\
				12	0.2628\\
				15	0.157255\\
				18	0.088675\\
				21	0.047175\\
				24	0.0239166666666667\\
				27	0.012512\\
			};
			\addlegendentry{Pilot-based scheme}
			
			\addplot [color=red, line width=1.5pt, mark size=5pt, mark=triangle, mark options={solid, red}, forget plot]
			table[row sep=crcr]{%
				0	0.9118\\
				3	0.85635\\
				6	0.765\\
				9	0.6386\\
				12	0.49495\\
				15	0.336055\\
				18	0.210965\\
				21	0.12055\\
				24	0.06584\\
				27	0.03427\\
			};
			
			\addplot [color=blue, dashdotted, line width=1.5pt, mark size=4.0pt, mark=o, mark options={solid, blue}, forget plot]
			table[row sep=crcr]{%
				0	0.94225\\
				3	0.89255\\
				6	0.81845\\
				9	0.7081\\
				12	0.56225\\
				15	0.39793\\
				18	0.25737\\
				21	0.15259\\
				24	0.08414\\
				27	0.044932\\
			};
			
			\addplot [color=green, dashed, line width=1.5pt, mark size=3.5pt, mark=square, mark options={solid, green}, forget plot]
			table[row sep=crcr]{%
				0	0.90445\\
				3	0.8479\\
				6	0.74615\\
				9	0.625\\
				12	0.4782\\
				15	0.344425\\
				18	0.23212\\
				21	0.1479\\
				24	0.0880966666666667\\
				27	0.051072\\
			};
			
			\addplot [color=mycolor2, dotted, line width=1.5pt, mark size=5pt, mark=diamond, mark options={solid, mycolor2}, forget plot]
			table[row sep=crcr]{%
				3	0.844595\\
				6	0.749625\\
				9	0.622765\\
				12	0.472915\\
				15	0.3366755\\
				18	0.2260855\\
				21	0.1483915\\
				24  0.0899366666666667\\
				27  0.0549427142857143\\
			};
			
			\addplot [color=mycolor1, dotted, line width=1.5pt, mark size=5.0pt, mark=asterisk, mark options={solid, mycolor1}, forget plot]
			table[row sep=crcr]{%
				0	0.9021\\
				3	0.83225\\
				6	0.72345\\
				9	0.5767\\
				12	0.4115\\
				15	0.270335\\
				18	0.15953\\
				21	0.08781\\
				24	0.0463133333333333\\
				27	0.0244\\
			};
			
			\addplot [color=black, dashed, line width=1.5pt, mark size=5.0pt, mark=x, mark options={solid, black}, forget plot]
			table[row sep=crcr]{%
				0	0.91805\\
				3	0.866\\
				6	0.78305\\
				9	0.6714\\
				12	0.5226\\
				15	0.36668\\
				18	0.229895\\
				21	0.133435\\
				24	0.07228\\
				27	0.038268\\
			};
			
		\end{axis}
		
		\begin{axis}[%
			width=5.833in,
			height=4.375in,
			at={(0in,0in)},
			scale only axis,
			xmin=0,
			xmax=1,
			ymin=0,
			ymax=1,
			axis line style={draw=none},
			ticks=none,
			axis x line*=bottom,
			axis y line*=left
			]
			\draw [black] (axis cs:0.377286,0.704762) ellipse [x radius=0.0191429, y radius=0.0547619];
			\draw [black] (axis cs:0.482643,0.747619) ellipse [x radius=0.0191429, y radius=0.0547619];
			
			\node [] (A) at (0.3,0.58) {\Large$\eta = 2$};
			\node [] (B) at (0.38,0.66) {};
			
			\node [] (C) at (0.6,0.855) {\Large$\eta = 3$};
			\node [] (D) at (0.49,0.77) {};
			
			\path [-latex] (A) edge node[right] {} (B);
			\path [-latex] (C) edge node[left] {} (D);
			
		\end{axis}
	\end{tikzpicture}%
	
}
    \end{center}
    \caption{SER performance of different Grassmannian constellations and a pilot-based scheme for $T = 2$ symbol periods, $M = N = 1$ antenna and spectral efficiency $\eta = R / M \in \{2,3\}$ bits/s/Hz from \cite{Cuevas24constellationsOnTheSphere}.}
    \label{fig:SER_T2}
\end{figure}

As an example, Fig. \ref{fig:SER_T2} shows the uncoded \gls{SER} performance of several structured and unstructured Grassmannian constellations for $T = 2$ symbol periods and $M = N = 1$ antenna. We include in the plot the structured Grass-Lattice \cite{Cuevas24constellationsOnTheSphere} (with low-complexity decoder), Cube-Split \cite{Ngo_cubesplit_journal} (with low-complexity decoder), Exp-Map \cite{Kammoun2007noncoherentCodes} (with low-complexity decoder), and unitary uniquely factorable constellations (UUFC) \cite{LiDesign2019} (with \gls{ML} decoder), as well as the unstructured Grassmannian constellations proposed in~\cite{Cuevas23unionBound} (with ML decoder), which are labeled as UB-Opt. In addition, we include the performance of a coherent pilot-based scheme, where the first symbol of the coherence interval is the pilot, which is known at the receiver, and the second symbol is taken from a QAM constellation with cardinality such that the coherent scheme has the same spectral efficiency as noncoherent schemes. For the comparison to be fair, the QAM constellations are normalized such that the average transmit power of the pilot-based scheme is the same as that of the noncoherent schemes and the power devoted to the data transmission is the same as the power devoted to training (optimal power allocation for $T = 2$ and $M = 1$ \cite{Hassibi2003howmuchtraining}).

We can observe that structured constellations perform slightly worse than the unstructured UB-Opt constellation in terms of \gls{SER}, as it was expected. Notice that UB-Opt uses the optimal \gls{ML} detector, whereas structured constellations uses a suboptimal detector with much lower complexity. Besides, Grassmannian constellations show similar or even better error rates than coherent pilot-based schemes.

\subsection{Multi-User Constellation Design}

Grassmannian constellations have also been used for multi-user scenarios with multiple transmitters (multiple access channels) or receivers (broadcast channels).

\subsubsection{Noncoherent Multiple Access\note{- Hoang and Diego}}

    
    
    

    Consider $K$ users, each with $M$ antennas, transmitting simultaneously to the receiver with the same power $P$. 
    The received signal over a coherence interval is 
    \begin{equation}
        \Ym = \sum_{k=1}^K \Xm_k \Hm_k + \sqrt{\frac{M}{TP}} \Zm
    \end{equation}
    where $\Xm_k \in \CC^{T \times M}$ and $\Hm_k \in \CC^{M \times N}$ are the transmitted signal and channel matrix of user~$k$, subject to the same assumptions as in the single-user case. User~$k$ has constellation $\Xc_k$ with cardinality $|\Xc_k| = L_k$. We are concerned with the design of the joint constellation $\Xc = \Xc_1 \times \dots \times \Xc_K$. Different from the single-user case, the optimal signaling for the multi-user case is not well-understood. Limited results have been reported on the capacity region of the noncoherent block-fading \gls{MAC}: a conjecture that the sum capacity can be achieved by no more than $T$ users~\cite{Shamai2002multiuser}, an achievable \gls{DoF} region for the two-user \gls{MIMO} case~\cite{Utkovski2013prelog_MAC}, and the optimal \gls{DoF} region for the two-user \gls{SIMO} case~\cite{Ngo2018DoFMAC}. However, these theoretical results do not provide clear insights into the structure of the optimal input. Letting each user employ \gls{USTM} independently from other users entails a small loss in terms of sum capacity~\cite{Devassy2015}. 

    Several existing joint constellation designs leverage the single-user Grassmannian designs: precoding individual Grassmannian constellations of lower dimension~\cite{NgoAsilomar2018multipleAccess}, using independently optimized single-user constellations~\cite{Li2021GrassmannianCodebook}, and optimizing a large single-user constellation of size $L_1 + \dots + L_K$ that is then partitioned according to some subspace distance measures into $K$ single-user constellations~\cite{Ngo2022joint_constellation_design}.
    
    More tailored design criteria for the \gls{MAC} are based on analyzing the \gls{PEP} of the joint symbols. The asymptotic expression of the single-user \gls{PEP} for the optimal receiver~\cite[Prop.~4]{Brehler2001asymptotic} is extended to the multi-user joint \gls{PEP}~\cite{Brehler2001noncoherent}. This extension is valid when the full diversity of $NM$ for each user is achieved, for which the coherence time must satisfy $T \ge (K+1)M$. Similar to the single-user case, this results in a design criterion consisting in maximizing the sum of $\det^{-N} (\Id_M - \Xm^\H \Xm' \Xm^\H \Xm')$ over all pairs of joint symbols $\Xm = [\Xm_1 \ \dots \ \Xm_K]$ and $\Xm' = [\Xm'_1 \ \dots \ \Xm'_K]$. In~\cite{Ngo2022joint_constellation_design}, some proxies for the \gls{PEP} error exponents between two joint symbols $\Xm$ and $\Xm'$ are proposed. These proxies have a natural geometric interpretation as the Riemannian distance between $\Id_T + \Xm \Xm^\H$ and $\Id_T + \Xm' {\Xm'}^\H$. They are valid also in the non-full diversity case where $T < (K+1)M$.
    Riemannian optimization techniques have been developed in~\cite{Alvarez2023constrained} to numerically solve the joint constellation design criteria proposed in~\cite{Brehler2001noncoherent} and~\cite{Ngo2022joint_constellation_design}. Since the optimality of \gls{USTM} or Grassmannian constellations is no longer valid in the multi-user case, \cite{Alvarez2023constrained} not only proposes to optimize the PEP expression in \cite{Brehler2001noncoherent} and its bounds or proxies in \cite{Ngo2022joint_constellation_design} on the Grassmann manifold but also considers two other different Riemannian manifolds: the \textit{oblique manifold}, which considers a per-codeword power constraint, and the \textit{trace manifold}, which uses an average power constraint. The manifold on which the optimization is performed can have a significant impact on error rate performance, especially in non-full diversity scenarios. The results in \cite{Alvarez2023constrained} suggest that, in the non-full diversity case, the proxy functions proposed in \cite{Ngo2022joint_constellation_design} optimized on the trace manifold are the best performing designs, whereas in the full diversity case the best performing constellations are those designed by optimizing the PEP expression in \cite{Brehler2001noncoherent} on the Grassmann manifold.


    The mentioned designs aim to optimize the joint constellation when an optimal \gls{ML} detector is deployed. In practice, a low-complexity (although suboptimal) detector would be preferred. For the precoding-based design in~\cite{NgoAsilomar2018multipleAccess}, the structure of the precoders allows to reduce the joint multi-user detection into $K$ single-user detection problems. For general constellations, \cite{Ngo2020multiuser_EP} proposed an efficient soft detector to compute the posterior marginal \gls{PMF} of the per-user symbols for the \gls{SIMO} \gls{MAC}. This detector is based on expectation propagation. Another approach that uses deep learning techniques has been proposed in \cite{Xue21}, where an end-to-end learning approach is used for joint transmitter and noncoherent receiver design on both \gls{iid} Rayleigh and spatially correlated channels.
    
\subsubsection{Noncoherent Broadcast\note{- Hoang}}
    
    For the \gls{BC}, the transmitter sends a superposition of the signals carrying individual information to each user and possibly common information to all users. Let us review the main superposition techniques via an example of two users. 
    
    The first superposition technique is 
    \textit{rate splitting}~\cite{Mao2022_rateSplitting}, where the transmitted signal is $\Xm =  \Xm_1 \Vm_1 +  \Xm_2 \Vm_2 +  \Xm_0 \Vm_0$, where $\Xm_k$, $k\in \{1,2\}$, contains the private information for user~$k$, $\Xm_0$ carries common information, and $\Vm_k,\Vm_0$ are precoders.  For precoded downlink transmission, rate splitting was originally used for the setting with perfect \gls{CSI} at the receiver and perfect/partial \gls{CSI} at the transmitter~\cite{Dai2016_rateSplitting,Joudeh2016_rateSplitting,Li2020_rateSplitting}. It was later designed for the noncoherent setting with only statistical \gls{CSI} in~\cite{Zhang2022_transmitCorrelationDiversity}. By exploiting transmit correlation diversity, i.e., the difference between the spatial correlation observed by different users, the authors design the precoders such that user~$k$ can only observe $\Xm_k$ and $\Xm_0$. 
    
    Another novel approach of rate splitting was recently proposed in \cite{Kancharana2024SparseRegressionCodesNoncoherent}, where sparse regression codes (SPARCs) are used in absence of \gls{CSI} at both transmitter and receiver. Assuming a base station with $M$ antennas that sends messages to $K$ single-antenna users in a noncoherent manner, the joint codebook $\Xm =[\Xm_1 \ | \ \Xm_2 \ | \ \ldots \ | \ \Xm_K]$, which is composed of individual codebooks $\Xm_k$, $k=1,\ldots,K$, is optimized to minimize the maximum coherence between columns of $\Xm$. The base station sends a superposition $\xv = \Xm  \sv$ where $\sv$ is a $K$-sparse vector with exactly $K$ non-zero entries corresponding to the locations of the chosen columns. Sparse signal recovery techniques can then be applied at the receiver to recover the best linear combination of the columns of $\Xm$ that approximates a vector of observations $\yv = h \xv+ \nv$.
    

    Another type of superposition is \textit{product superposition}, first proposed for the setting with a static user with perfect \gls{CSI} (say, user~$1$), and a dynamic receiver with only statistical \gls{CSI} (user $2$)~\cite{Li2012productsuperposition}. The transmitted signal is $\Xm = \Xm_2 \Xm_1$. Grassmannian signal is used to construct $\Xm_2$, so that information is carried out by its column space, which is invariant if $\Xm_1$ is full rank. The static user decodes and removes $\Xm_2$ from the received signal, then decodes $\Xm_1$. A pilot-based strategy can also be used to design $\Xm_2$~\cite{Li2015coherent_product_superposition}, which is adopted for the noncoherent setting with only statistical \gls{CSI} in~\cite{Zhang2022_transmitCorrelationDiversity}. Notably, \cite{Zhang2022_transmitCorrelationDiversity} shows that, with transmit correlation diversity, a combination of rate splitting and product superposition can significantly improve the achievable \gls{DoF} and rate regions compared to orthogonal resource sharing. 

    A novel form of product superposition is \textit{Grassmannian superposition}~\cite{Schwarz2021noncoherent_broadcasting}, where individual data streams are carried in nested subspaces. 
    Formally, a sequence of nested subspaces defines another manifold called flag manifold \cite{ye2022optimization,FlagsCVPR2025}, with a particular geometry that can be exploited for designing constellations in the flag and for their efficient detection. For the \gls{SIMO} case, the transmitted signal is $\xv = \Qm_1 \times \dots \times \Qm_B$ where $B$ is the number of streams and $\Qm_b$ represents a point in $\GG(d_{b}, \CC^{d_{b-1}})$ with $d_0 = T$ and $d_B = 1$. This scheme is suitable for multi-resolution transmission. The design of Grassmannian constellations for each nested subspace was further investigated in~\cite{Schwarz2021noncoherent_multi_resolution,Schwarz22codebookTraining}, and a soft successive detection scheme was proposed in~\cite{Schwarz23approximate}.

Table \ref{tab:grassmann_designs_multiuser} summarizes the key contributions of the previously mentioned works on multi-user Grassmannian constellation design.

\begin{table*}[t!]
\renewcommand{\arraystretch}{1.1}
\caption{Literature overview on multi-user Grassmannian constellation design}
\centering
\begin{tabular}{ |m{3.5cm}|m{.7cm}|m{1.5cm}|m{10.5cm}| } 
\hline
\textbf{Design} & \textbf{Year} & \textbf{Scenario} & \textbf{Main idea} \\
\hline \hline
Ngo \textit{et al.} \cite{NgoAsilomar2018multipleAccess} & 2018 & SIMO MAC & Precoding individual Grassmannian constellations of lower dimension\\
\hline 
Schwarz \textit{et al.} \cite{Schwarz2021noncoherent_multi_resolution} & 2021 & SIMO BC & Grassmannian superposition using nested subspaces\\
\hline
Xue \textit{et al.} \cite{Xue21} & 2021 & SIMO MAC & End-to-end learning approach for joint transmitter and noncoherent receiver design \\
\hline
Li \textit{et al.} \cite{Li2021GrassmannianCodebook} & 2021 & SIMO MAC & Independently optimized single-user constellations + short spreading sequence as signature of multiple access\\
\hline
Schwarz \textit{et al.} \cite{Schwarz22codebookTraining} & 2022 & MIMO BC & Trellis-based hierarchichal Grassmannian classification for nested subspaces\\
\hline
Zhang \textit{et al.} \cite{Zhang2022_transmitCorrelationDiversity} & 2022 & MIMO BC & Additive rate splitting and product superposition for noncoherent setting with only statistical \gls{CSI}\\
\hline
Ngo \textit{et al.} \cite{Ngo2022joint_constellation_design} & 2022 & MIMO MAC & Optimizing a large single-user constellation that is then partitioned according to some subspace distance measures into several single-user constellations\\
\hline
Álvarez-Vizoso \textit{et al.} \cite{Alvarez2023constrained} & 2023 & MIMO MAC & Riemannian optimization techniques to numerically solve the joint constellation design criteria proposed in~\cite{Brehler2001noncoherent} and~\cite{Ngo2022joint_constellation_design}\\
\hline
Kancharana \textit{et al.} \cite{Kancharana2024SparseRegressionCodesNoncoherent} & 2024 & SIMO BC & Use of sparse regression codes (SPARCs)\\
\hline
\end{tabular}
\label{tab:grassmann_designs_multiuser}
\end{table*}


\section{Energy-Based Schemes} \label{sec:energy-based}
\note{Ana and Ruben}

Communication systems started being energy-based, e.g., amplitude-modulation radio, but they were rapidly overtaken by systems capable of using the phase due to their robustness against amplitude variations, which led to significantly better performance. One of the benefits brought by the arrival of massive MIMO was channel hardening, a statistical effect that decreases the amplitude variations of a communications channel if the number of antennas is large. This effect, which would solve the sensitivity of energy-based communication systems to amplitude variations of the received signal, 
has motivated research in their use in combination with massive MIMO.

\subsection{Theoretical Foundation}

Energy-based schemes were proposed in \cite{chowdhuryDesignPerformanceNoncoherent2014} (and extended in~\cite{Manolakos2016noncoh_Energybased_massiveSIMO}) as an alternative way of transmission and decoding of information to the standard coherent MIMO coding schemes for SIMO and the uplink of multi-user MIMO scenarios. They encode information in a way that does not require tracking fast channel variations, which avoids the complexity of instantaneous channel estimation.

Energy-based techniques use underlying channel statistics to build an asymptotically stable decoding scheme. Therefore, these schemes are indifferent to instantaneous variations in the channel and work for block fading and smoothly varying channel models. As long as the underlying channel statistics do not change, these schemes are viable even when the coherence time is equal to the symbol time~\cite{chowdhuryDesignPerformanceNoncoherent2014}. 

The main idea of these schemes is to encode the information in the amplitude (or power) of the signal. Then, due to the law of large numbers, the average received energy along the antennas of a large MIMO array will converge to the energy of the sent signal minus the free-space path loss (expressed in dB). If the free space path loss is known or compensated, then a large number of spatially uncorrelated channels (antennas) can be used to diminish the effects of fading. This is true for Rayleigh and Ricean fading channels~\cite{chowdhuryDesignPerformanceNoncoherent2014}. The proposed demodulation scheme consists of taking the average energy of the received symbol across all the antennas and mapping it into a 1-dimensional constellation. The received symbol follows the SIMO system model 
\begin{equation} \label{eq:Energy_based_detection1}
    \yv = x \hv  + \wv,
\end{equation}
where $x \in \CC$ is the transmitted symbol, $\wv \in \CC^{1 \times \Nr}$ is the \gls{AWGN} vector, $\hv \in \CC^{1 \times \Nr}$ is the \gls{SIMO} channel vector, $\yv \in \CC^{1 \times \Nr}$ is the received symbol vector, and $\Nr$ is the number of receive antennas. The average energy of the received symbol across the $N$ receiving antennas is obtained as 
\begin{equation} \label{eq:Energy_based_detection2}
    \frac{1}{N}\|\yv\|^2 = \frac{1}{N} \sum_{n=1}^{N} |y_n|^2,
\end{equation}
and is then mapped into the used $1$-dimensional constellation. Because the decoding regions are related to the second-order channel statistics, these schemes require those to be stable, as well as good power control and spatially uncorrelated channels. Not meeting these requirements will hinder the performance. 

Energy-based schemes excel in the asymptotic regime, where the number of receiving antennas is large. Under the previously exposed assumptions, and in the narrowband case, they can get the same performance scaling (in terms of BER and SNR as functions of the number of receive antennas) 
as coherent systems~\cite{Manolakos2016noncoh_Energybased_massiveSIMO, manolakosCSINotNeeded2014,chowdhuryScalingLawsNoncoherent2016}. In~\cite{gomez-cubaCapacityScalingNonCoherent2019}, it is shown that energy-based systems can also work in wideband systems but with worse asymptotic scaling. 

\subsection{Single-User Constellation Design}
The design of constellations in these schemes follows the same principles as in coherent schemes, i.e, to minimize the distance between the different symbols. 
Due to the constellation only having one dimension because of the energy being a scalar, the transmitted symbol has the only constraint of its energy belonging to the levels defined in the constellation, i.e., the symbol must be chosen from
\begin{align}
\Xc \defeq \big\{x_1,\dots,x_{L} \in \CC \colon |x_i|^2 = p_i, i\in [L] \big\}.
\label{eq:energy_constellation}
\end{align}
Due to this constraint, the constellation design problem quickly becomes cumbersome, especially if the knowledge of the channel statistics is not perfect~\cite{manolakosConstellationDesignNoncoherent2014}. 
As a result, designing constellations for energy-based systems is an NP-hard optimization problem~\cite{chowdhuryScalingLawsNoncoherent2016, manolakosCSINotNeeded2014}. The work~\cite{manolakosConstellationDesignNoncoherent2014} presents a robust constellation design scheme that accounts for the possible uncertainty in the knowledge of channel statistics. Later, \cite{hanConstellationDesignEnergyBased2022} presents a constellation design approach suitable for correlated channels and compares it to the optimal constellation for uncorrelated channels from~\cite{Vucetic19}. Finally, \cite{martiConstellationDesignQuadratic2024} presents a lower complexity constellation design approach suitable for quadratic detectors.

\subsection{Multi-User Constellation Design}
\begin{figure}[t!]
    \centering
    \includegraphics[width=\linewidth]{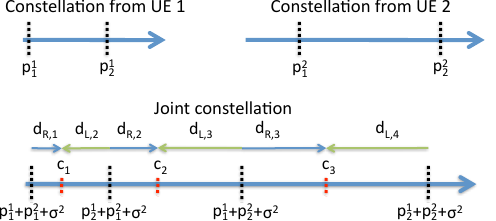}
    \caption{Example of multi-user energy-based constellation adapted from the single user constellation from~\cite{manolakosConstellationDesignNoncoherent2014}. Each region corresponds to a predefined combination of symbols from each user.}
    \label{fig:energy_constellation}
\end{figure}

Regarding the multi-user case~\cite{manolakosCSINotNeeded2014}, the multiplexing capabilities of the mentioned schemes are very limited. This happens due to their inability to multiplex users spatially. As a result, the only way to multiplex users is in the one-dimensional constellation domain, by assigning certain decoding regions to each combination of user symbols, see Fig.~\ref{fig:energy_constellation}. Like in the Grassmannian case, this is called the joint constellation, and it tightens the requirement of knowledge of channel statistics. 

In Fig.~\ref{fig:energy_constellation}, $c_{1}, c_2, c_3$ represent the thresholds used in the receiver to assign the received signal to its corresponding decoding region, for the two user case. The constellation design process aims at defining the optimal thresholds that minimize the SER. The probabilities of error are determined by the distances $d_{L/R, i}$ separating each decoding region and the adjacent thresholds. In the multi-user case, the statistics of each of the users' channels must be known and accounted for in the constellation design process~\cite{manolakosCSINotNeeded2014}.

For clarity and ease of comparison, Table \ref{tab:differential_papers} summarizes the previously discussed works on energy-based schemes, emphasizing their key contributions.

\begin{table*}[ht]
\renewcommand{\arraystretch}{1.1}
\centering
\caption{Literature overview on energy-based schemes}
    \begin{tabular}{ |m{3.5cm}|m{.7cm}|m{2cm}|m{10cm}| } 
        \hline
        \textbf{Design} & \textbf{Year} & \textbf{Scenario} & \textbf{Main idea} \\
        \hline \hline
        Chowdhury \textit{et al.}~\cite{chowdhuryDesignPerformanceNoncoherent2014} & 2014 & single-user SIMO & MIMO energy-based schemes \\
        \hline
        Manolakos \textit{et al.}~\cite{manolakosConstellationDesignNoncoherent2014} & 2014 & single-user SIMO & single user constellation design \\
        \hline
        Manolakos \textit{et al.}~\cite{manolakosCSINotNeeded2014} & 2014 & multi-user MIMO & Multi-user constellation design \\
        \hline
        Manolakos \textit{et al.}~\cite{Manolakos2016noncoh_Energybased_massiveSIMO} & 2016 & single and multi-user SIMO & Extension of~\cite{chowdhuryDesignPerformanceNoncoherent2014, manolakosConstellationDesignNoncoherent2014} and~\cite{manolakosCSINotNeeded2014}, and study of practical implementations \\
        \hline
        Gómez-Cuba \textit{et al.}~\cite{gomez-cubaCapacityScalingNonCoherent2019} & 2019 & single-user SIMO & Extension to wideband channels with formal derivations of the capacity for different scenarios \\
        \hline
        Han \textit{et al.}~\cite{hanConstellationDesignEnergyBased2022} & 2022 & single-user SIMO & Constellation design for correlated channels \\
        \hline
        A. Mart\'{i} \textit{et al.}~\cite{martiConstellationDesignQuadratic2024} & 2024 & single-user SIMO & Low complexity constellation design for quadratic detectors \\
        \hline
    \end{tabular}
\label{tab:energy_papers}
\end{table*}

\section{Schemes based on Differential Detection} \label{sec:differential}
\note{Ana and Ruben}

Differential schemes encode information using phase differences, providing better performance and increased multi-user multiplexing capabilities with respect to energy-based schemes. A representative example is Digital Audio Broadcasting, used since the late 1990s with noncoherent differential phase detection. Recently, massive numbers of antennas bring attractive capabilities that could greatly enhance the performance of these schemes and have motivated additional research.

\subsection{Theoretical Foundation}
 Differential schemes are built on the assumption of channel changes being smooth. The main idea is to encode each information symbol as an added phase shift to the previously encoded one, so that 
\begin{equation} 
    x[n] = x[n-1]s[n], \quad n > 1,
\end{equation}
where \(x[n]\) corresponds to the unit-modulus differentially encoded symbol at time \(n\), and \(s[n]\) to the unit-modulus original information symbol. The demodulation is done by multiplying each received symbol \(y[n]\) by the complex conjugate of the previously received one \(y[n-1]\), as
\begin{equation} \label{eq:diff_demod}
    z[n] = y[n-1]^* y[n].
\end{equation}
where $z[n]$ is the demodulated received symbol that is mapped to the used PSK constellation through an \gls{ML} (or minimum distance) detector. On the one hand, because the information is encoded in the phase, the requirement of invariability of the second-order channel statistics from energy-based systems is relaxed for the single-user case. On the other hand, as the information is encoded in the phase differences between two symbols, there is always a first symbol $x[1]$ that must be transmitted without carrying any information and constitutes a small overhead. As a result, these systems are not advisable for short, isolated, discontinuous transmissions, which are modeled through block-fading channels.  

The promising asymptotic scaling of MIMO systems motivated the combination of these schemes with massive numbers of antennas. The first appearance of differential schemes combined with massive MIMO systems was in~\cite{schenkNoncoherentDetectionMassive2013}, which uses a very restrictive channel model based on an Ultra Wide Band scenario. The work in \cite{armadaNoncoherentMultiuserLarge2015} further developed noncoherent differential massive MIMO systems to Rayleigh fading channel models, as well as proposed multiplexing in the constellation domain. The way these systems make use of the high numbers of antennas is by averaging the scalar product of the received symbol \(y_i[n]\) and the previously received one \(y_i[n-1]\) over the receive antennas 
\begin{equation} \label{eq:diff_demod_averaging}
    z[n] = \frac{1}{N} \sum_{i=1}^{N} y_i[n-1]^* y_i[n],
\end{equation}
where \(N\) is the number of receive antennas.
Averaging over the receive antennas exploits the channel hardening effect and diminishes the impact of noise. The received constellation \(z[n]\) is a sum of the constellation of the user, plus some noise and interference terms that decrease as the number of antennas increases~\cite{armadaNoncoherentMultiuserLarge2015}. These schemes are designed to work under smoothly varying channel conditions. Due to their use of the previous symbol to adaptively remove the channel effect on the phase, they need channel variations to be negligible during two consecutive symbol periods. Moreover, for channel hardening to be effective, they require the channels from the different antennas to be uncorrelated~\cite{moralesEffectSpatialCorrelation2021}. 
Finally, in the multi-user case, because of their inability to spatially multiplex users, these schemes need stable power control and knowledge of the second-order channel statistics, so that the multi-user joint constellation maintains its shape. 

Like energy-based schemes, differential-detection-based schemes also benefit from asymptotically large numbers of antennas, getting similar capacity gains to coherent systems~\cite{armadaNoncoherentMultiuserLarge2015}. Under extremely short coherence times, differential-detection-based schemes have been shown to outperform their coherent counterparts~\cite{chenhuNonCoherentMassiveMIMOOFDM2020}.

We next present generalizations of DPSK that include the antenna dimension of MIMO communications. As it will be discussed in the incoming sections, these schemes are more adequate for MIMO setups with moderate numbers of antennas.

\subsubsection{Differential USTM} \label{DUSTM}
Differential Unitary Space-Time Modulation (DUSTM) was introduced in~\cite{hughesDifferentialSpacetimeModulation2000, HochwaldDiff00} and is the generalization of DPSK schemes for multiple antennas along several channel uses. They were initially proposed as a way to exploit spatial diversity in MIMO scenarios, maximizing the achievable rate, while not having to estimate the channel. 

The principle of differential encoding of the information is still present, but instead of transmitting single DPSK symbols,  the transmitter sends USTM symbols, 
i.e., each symbol spans all the transmit antennas $M$ during $M$ channel uses~\cite{Cabrejas2016non_coherent} as
\begin{equation} \label{eq:DUSTM}
    \Xm_{\tau} = \Sm_{\tau} \Xm_{\tau-1},
\end{equation}
where $\Xm_{\tau} \in \CC^{\Nt \times \Nt}$ is the currently transmitted USTM symbol, $\Xm_{\tau-1} \in \CC^{\Nt \times \Nt}$ is the previously transmitted DUSTM symbol, and $\Sm \in \CC^{\Nt \times \Nt}$ is the original DUSTM symbol mapped from information bits and taking values in the constellation $\{\Sm_1,\dots,\Sm_L\}$. For DUSTM to work, the effect of the channel must be similar between two adjacent USTM symbols, i.e, during $2M$ channel uses. This causes the differential symbol $\Sm_l$, which is encoded between two consecutive received USTM symbols $\big\{ \Ym_{\tau-1} , \Ym_{\tau-1} \big\} \in \CC^{\Nt \times N} $, to be unchanged, so that  
\begin{align} \label{eq:DUSTM_tx_1}
    \Ym_{\tau-1} & = \Xm_{\tau-1} \Hm + \Wm_{\tau-1}, \\
    \Ym_{\tau} & = \Xm_{\tau} \Hm + \Wm_{\tau}, \label{eq:DUSTM_tx_2}    
\end{align}
where $\Wm \in \CC^{\Nt \times N}$ is a matrix of additive independent $\Cc\Nc(0, \sigma^2)$ noise, with $\sigma^2$ being the noise variance, and $\Hm \in \CC^{\Nt \times N}$ is the channel matrix,  
which is constant during the transmission of two consecutive DUSTM symbols, i.e., during $2M$ channel uses. The differential demodulation from~\eqref{eq:diff_demod} is still valid in matrix form, which allows the detection to be done through an ML decoder as
\begin{equation}
\hat{\Sm} = \Sm_{\hat \ell}, \quad \hat{\ell} = \arg\min_{\ell} \| \Ym_{\tau} - \Sm_{\ell}\Ym_{\tau-1}\|,
\end{equation}
where $\mathbf{\hat{\Sm}}$ is the ML decoded DUSTM symbol. Due to the correlator having the noise power of two different symbols as input, the noise variance of these scheme doubles with respect to coherent ones, so that
\begin{align}
\begin{split}
\Ym_{\tau} &= \Sm_{\tau} \Ym_{\tau-1} + \Wm_{\tau} - \Sm_{\tau} \Wm_{\tau-1} \\
 &= \Sm_{\tau} \Ym_{\tau-1} + \sqrt{2} \, \Wm'_\tau,
\end{split}
\end{align}
where the last step uses the fact that the noise matrices $\Wm_{\tau}$ and $\Wm_{\tau-1}$  
are independent and statistically invariant to multiplication by unitary matrices to obtain $\Wm'_\tau  \in \CC^{\Nt \times N} $, an additive independent $\Cc\Nc(0, \sigma^2)$ noise matrix. This results in a $3$~dB performance loss with respect to coherent schemes with the same noise variance. 

DUSTM schemes, like Grassmannian designs, exhibit poor scalability with increasing numbers of antennas. The fundamental issue arises from the fact that each additional antenna introduces a new dimension (i.e., an additional row) to the DUSTM symbol, leading to considerable complexity in constellation design as the number of antennas becomes large (e.g., on the order of 100 antennas). This scaling challenge limits the practical applicability of DUSTM in massive MIMO scenarios.

\subsubsection{Differential \gls{STBC}} \label{DSTBC}

These schemes are a differential extension of the \gls{STBC}, namely the Alamouti code~\cite{Alamouti98}. The main idea is to send differentially encoded symbols in the orthogonal space-time streams of the Alamouti code~\cite{tarokhDifferentialDetectionScheme2000}. At the receiver, the effect of the channel can be eliminated by multiplying by the complex conjugate of the previously received space-time stream. Therefore, the need for channel estimates is eliminated while still maintaining the attractive spatial diversity properties at rate one of the Alamouti codes. 

The proposal of these schemes is to transmit differentially encoded symbols using the Alamouti scheme. The symbols $s_1$ and $s_2$ will be differentially encoded in the space-time bases transmitted by the Alamouti scheme. The symbol $s_1$ is differentially encoded into the $(x_{2t-1}, \; x_{2t})$ stream, and $s_2$ is encoded into the $(-x_{2t}^*, \; x_{2t-1}^*)$ stream
\begin{equation}
(x_{2t+1}, \; x_{2t+2}) = s_1 (x_{2t-1}, \; x_{2t}) + s_2 (-x_{2t}^*, \; x_{2t-1}^*).
\end{equation}

Consider a transmitter with two antennas. During one channel use, antennas one and two transmit $(x_{2t-1}, \; x_{2t})$, and during the next one they transmit $(-x_{2t}^*, \; x_{2t-1}^*)$. The decoding is done by applying the complex conjugate multiplication from the Alamouti scheme, but instead of using the channel estimate for the multiplication, the previously received differentially encoded symbols are used. Due to the differential Alamouti encoding, by multiplying by the previously received space-time sequences $(y_{2t-1}, \; y_{2t})$ and $(-y_{2t}^*, \; y_{2t-1}^*)$, the channel effect can be reduced to two orthogonal space-time streams multiplied by the sum of the squared channel magnitudes
\begin{equation}
\begin{split}
y_1 &= y_{2t+1} y_{2t-1}^* + y_{2t+2} y_{2t}^* \\
&= (|h_{1,1}|^2 + |h_{2,2}|^2) s_1 + n_1,
\end{split}
\end{equation}
for the symbol $s_1$, which is transmitted in the $(x_{2t-1}, \; x_{2t})$ space time stream. And
\begin{equation}
\begin{split}
y_2 &= y_{2t+1} y_{2t}^* - y_{2t+2} y_{2t-1}^* \\ 
&= (|h_{1,1}|^2 + |h_{2,2}|^2) s_2 + n_2, 
\end{split}
\end{equation}
for the symbol $s_2$, which is transmitted in the $(-x_{2t}^*, \; x_{2t-1}^*)$ stream. Here, $h_{j,i}$ represents the channel coefficient from the transmitter antenna $j$ to the receiver antenna $i$, and $n_{l}$ represents the noise for the orthogonal space-time stream $l$. Once $y_1$ and $y_2$ are computed, they are mapped to the used PSK constellation through an ML detector.


The scalability and multi-user capabilities of these schemes are similar to those of the original Alamouti code. Like \glspl{STBC}, these schemes scale poorly with increasing numbers of antennas, which limits their use to traditional small-size MIMO setups. Consequently, they are not a good solution for multi-user MIMO applications involving high numbers of users.

\subsection{Single-User Constellation Design}

Designing constellations for differential-detection-based schemes also consists of maximizing the distance between symbols. But in this case, due to the symbols being encoded with reference to the previous one, the constellation must form a group under matrix multiplication, i.e., they must have internal composition, associativity, and have an identity and inverse element~\cite{HochwaldDiff00}. Internal composition means the matrix product of two group members must always be equal to another group member, e.g., PSK constellations. Internal composition also implicitly defines the inverse element. Associativity is inherent in the matrix multiplication operation. Finally, the identity element will be the identity matrix. As a result, the symbols of the constellation will be drawn from 
 \begin{align}
\Xc \defeq \big\{\Xm_1,\dots,\Xm_{L} \in \CC^{T\times M} \colon \Xm_i^\H\Xm_j = \Xm_k,  \{i,j,k \}\in [L] \big\}.
\label{eq:differential_constellation}
\end{align}
For the SIMO (or multi-user MIMO) case, the previous conditions reduce to the use of PSK contellations~\cite{armadaNoncoherentMultiuserLarge2015}. As an exception, \cite{kongDifferentialQAMDetection2016} extends the differential encoding to 16-APSK and 16-QAM constellations for the single user case; 16-APSK constellations also use differential encoding in the amplitude, but 16-QAM requires the use of a lookup table, which becomes cumbersome and limits the extension to higher order QAM schemes. Differential PSK-based schemes have been investigated for satellite applications and configurations, where the absence of CSI can be advantageous in ground-to-space links characterized by high Doppler components~\cite{ICSSC22}. Moreover, these schemes have also been proposed in spatial modulation configurations to reduce the required radio-frequency chains~\cite{MonzonWCL24}.

Moreover, in the MIMO case, where DUSTM schemes excel,~\cite{HochwaldDiff00} imposes the condition of the constellation forming an Abelian group, i.e., the group operation (matrix multiplication in this case) must be commutative. This is done to make the DUSTM symbol matrices diagonal and eliminate the need for a lookup table in the transmitter~\cite{HochwaldDiff00}.

Finally, differential STBC schemes, despite being used in a MIMO scenario, are affected by the limitations of SIMO schemes, i.e., having to use symbols belonging to PSK constellations~\cite{tarokhDifferentialDetectionScheme2000}. The work~\cite{hwangDifferentialSpaceTime2003} extends these schemes to work with QAM constellations but requires passive estimation of the amplitude of the channel coefficients, which works well only under high SNRs and long enough coherence times.

\subsection{Multi-User Constellation Design}
\begin{figure}[t!]
    \centering
    \includegraphics[width=\linewidth]{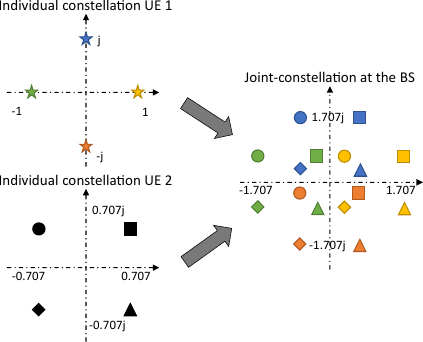}
    \caption{Example of differential joint constellation for two users from~\cite{chenhuNonCoherentMassiveMIMOOFDM2020}. The receiver first maps the received symbol to a point in the joint constellation, and then maps it to a pair of points, one from the first individual constellation (the color) and one from the second individual constellation (the shape). 
    } 
    \label{fig:differential_constellation}
\end{figure}

In the multi-user case, if there exists knowledge about the long-term channel statistics, i.e., the angles of the line-of-sight path from the users to the antenna, it is possible to separate them spatially~\cite{sternNonCoherentMultiUserMassive2021}. If there is no previous knowledge about the channel, then a joint constellation analogous to the one from energy-based schemes must be used; 
an example is shown in Fig.~\ref{fig:differential_constellation}.   
If the multiplexing is done in the constellation domain, the limitation to PSK individual constellations complicates the design of joint constellations for multiple users. In \cite{lopez-moralesConstellationDesignMultiuser2022}, the use of Monte-Carlo methods is proposed to solve the multi-user constellation design problem.  Moreover, constellation multiplexing requires stable power control so that the joint constellation maintains its shape. Moreover, the different users' channels must have low spatial correlation for the averaging between antennas \eqref{eq:diff_demod_averaging} to exploit the channel hardening effect and give the desired joint constellation at the exit of the differential receiver. 
Expanding \eqref{eq:diff_demod_averaging}, 
the effective received signal for the multi-user case results in  
\begin{equation} \label{eq:diff_multi_user}
    z[n] = \frac{1}{N} \sum_{k=1}^{K} \sum_{i=1}^{N} |h_{ki}|^2 \beta_k s_k[n] +  \zeta[n],
\end{equation}
where $h_{ki}$ represents the channel from user $k$ to antenna $i$, $\beta_k$ is the received power of the signal of user $k$, the first term of \eqref{eq:diff_multi_user} is the received joint symbol affected by the channel, and $\zeta[n]$ is the interference from the other users plus the noise. 

To address the issue of inter-user interference, ~\cite{victorElecEng21} has proposed user clustering schemes that tailor the constellation design according to the fading conditions experienced by each user group. This approach extends multi-user constellation design to heterogeneous environments, enabling more adaptive and interference-resilient communication strategies.

To facilitate comparison, Table \ref{tab:differential_papers} presents a summary of the previously discussed works on differential detection-based schemes, highlighting their main contributions.

\begin{table*}[ht]
\renewcommand{\arraystretch}{1.1}
\caption{Literature overview on differential-detection-based schemes}
\centering
    \begin{tabular}{ |m{3.5cm}|m{.7cm}|m{2.1cm}|m{10cm}| } 
        \hline
        \textbf{Design} & \textbf{Year} & \textbf{Scenario} & \textbf{Main idea} \\
        \hline \hline
        Howchald and Sweldens~\cite{HochwaldDiff00}/ Hughes~\cite{hughesDifferentialSpacetimeModulation2000} & 2000 & single-user MIMO & Differential Unitary Space Time Modulation (DUSTM) \\
        \hline
        Tarokh and Jafarkhani~\cite{tarokhDifferentialDetectionScheme2000} & 2000 & single-user MIMO & Differential Space Time Block Codes (STBC) \\
        \hline
        Hwang \textit{et al.}~\cite{hwangDifferentialSpaceTime2003} & 2003 & single-user MIMO & Differential STBC with nonconstant modulus constellations (QAM) \\
        \hline
        Schenk and Fischer~\cite{schenkNoncoherentDetectionMassive2013} & 2013 & multi-user MIMO & DPSK constellations for massive multi-user MIMO, requires weighting matrices to multiplex users  \\
        \hline
        Armada and Hanzo~\cite{armadaNoncoherentMultiuserLarge2015} & 2015 & multi-user MIMO & Simpler detection based on the complex conjugate, introduces the joint constellation concept for differential schemes \\
        \hline
        Kong \textit{et al.}~\cite{kongDifferentialQAMDetection2016} & 2016 & single-user SIMO & Differential 16 QAM and 16 APSK \\
        \hline
        Chen Hu \textit{et al.}~\cite{chenhuNonCoherentMassiveMIMOOFDM2020} & 2020 & multi-user MIMO & DPSK coding in the OFDM subcarrier domain \\
        \hline
        Monzon-Baeza \textit{et al.}\cite{victorElecEng21} & 2021 & multi-user MIMO & User grouping schemes for differential noncoherent detection \\
        \hline
        Stern and Fischer~\cite{sternNonCoherentMultiUserMassive2021} & 2021 & multi-user MIMO & Uses spatial multiplexing along with noncoherent detection \\
        \hline
        Lopez-Morales \textit{et al.}\cite{lopez-moralesConstellationDesignMultiuser2022} & 2022 & multi-user MIMO & Joint constellation design using genetic algorithms \\
        \hline
        Monzon-Baeza \textit{et al.}\cite{MonzonWCL24} & 2024 & multi-user MIMO & Spatial modulation with noncoherent detection \\
        \hline
    \end{tabular}
\label{tab:differential_papers}
\end{table*}
\section{Practical Aspects and Limitations \note{-- Diego}} \label{sec:practical}

In this section, we conduct a comprehensive analysis of the practical considerations and inherent limitations associated with the three considered noncoherent communication schemes: Grassmannian, energy-based, and differential detection-based schemes. 
In general, noncoherent schemes are designed under a set of assumptions that we summarize for each approach in Table~\ref{tab:assumptions}. Applying a scheme to other scenarios leads to model mismatch and potential performance degradation.
Next, drawing on the summary presented in Table \ref{tab:practical_aspects}, we systematically examine key factors such as compatibility with OFDM architectures, sensitivity to DC components, robustness to \glspl{HWI}, spectral efficiency constraints, and scalability with respect to the number of users. Each of these aspects is discussed in dedicated subsections, with references to representative works provided where applicable. This analysis offers critical insights into the trade-offs and design considerations that must be addressed when deploying noncoherent schemes in practical wireless communication systems.

\begin{table*}[t!]
\renewcommand{\arraystretch}{1.1}
\caption{Assumptions made by each noncoherent design approach}
\centering
    \begin{tabular}{ |m{6cm}|m{11.15cm}| }
        \hline
        \textbf{Scheme} & \textbf{Assumptions} \\
        \hline \hline
        \multirow{3}{*}{Grassmannian signaling} & $\bullet$ Coherence time at least $2M$ channel uses\\
         & $\bullet$ i.i.d. Rayleigh fading channel coefficients (no spatial correlation between antennas)\\
         & $\bullet$ Narrowband signal assumption\\
        \hline
        \multirow{5}{*}{Energy-based schemes} & $\bullet$ Coherence time longer than the symbol period\\
         & $\bullet$ Known second-order channel statistics\\
         & $\bullet$ Accurate and stable power control\\
         & $\bullet$ Low spatial correlation between antennas\\
         & $\bullet$ Narrowband signal assumption\\
        \hline
        \multirow{5}{*}{Differential PSK for massive multi-user MIMO} & $\bullet$ Coherence time larger than twice the symbol period\\
        & $\bullet$ Smoothly-varying channel\\
        & $\bullet$ Low spatial correlation between antennas\\
        & $\bullet$ Narrowband signal assumption\\
        & $\bullet$ Accurate and stable power control (only in the multi-user case)\\
        \hline
        Differential USTM / Differential STBC & Similar to Grassmannian signaling 
        \\
        \hline
    \end{tabular}
\label{tab:assumptions}
\end{table*}

\begin{table*}[t!]
    \renewcommand{\arraystretch}{1.1}
    \caption{A list of practical implementation aspects and limitations that affect noncoherent communication schemes}
    \centering
    \begin{tabular}{|m{5cm}| >{\centering\arraybackslash}m{3.75cm} | >{\centering\arraybackslash}m{3.75cm} | >{\centering\arraybackslash}m{3.75cm} |} 
    \hline
    \textbf{Practical aspect / limitation} & \textbf{Grassmannian schemes} & \textbf{Energy-based schemes} & \textbf{Differential detection-based schemes} \\
    \hline \hline
    Implementation with OFDM & \cellcolor{green!20}\cmark~\cite{Ayadi2009OFDM,Fouad2015OFDM,fanjul2017experimental} & \cellcolor{green!20}\cmark~\cite{Fazeli2020Generalized} & \cellcolor{green!20}\cmark~\cite{chenhuNonCoherentMassiveMIMOOFDM2020}\\
    \hline
    DC component impact & \cellcolor{orange!20}Medium~\cite{CuevasThesis2024} & \cellcolor{green!20}Low & \cellcolor{green!20}Low\\
    \hline
    \gls{HWI}s impact & \cellcolor{red!20}High~\cite{Cuevas24harware} & \cellcolor{red!20}High~\cite{Leila2019Performance} & \cellcolor{red!20}High~\cite{Bucher2020HWIs}\\
    \hline
    \multirow{2}{*}{Spectral efficiency} &\cellcolor{green!20}Structured $\rightarrow$ High & \cellcolor{red!20}Low~\cite{gomez-cubaCapacityScalingNonCoherent2019} & \cellcolor{green!20}High\\ \
    & \cellcolor{red!20}Unstructured $\rightarrow$ Low~\cite{Cuevas23unionBound} & \cellcolor{red!20} & \cellcolor{green!20}\\
    \hline
    Scalability with number of users & \cellcolor{red!20}Poor~\cite{Alvarez2023constrained,Ngo2022joint_constellation_design,Schwarz22codebookTraining,Schwarz23approximate} & \cellcolor{red!20}Poor~\cite{manolakosConstellationDesignNoncoherent2014,manolakosCSINotNeeded2014} & \cellcolor{red!20}Poor~\cite{lopez-moralesConstellationDesignMultiuser2022}\\
    \hline
    \end{tabular}
    \label{tab:practical_aspects}
\end{table*}

\subsection{Practical Implementation and Integration with OFDM Systems}

The practical implementation of noncoherent communication schemes varies significantly across the three main approaches considered in this work. Among them, differential detection-based schemes are the most mature and have received extensive attention in the literature, with numerous studies addressing their implementation challenges, performance under realistic conditions and hardware feasibility. Notably, differential modulation has been incorporated into some communication standards, such as IEEE 802.11 \cite{ieee80211-2020}, which uses \gls{DBPSK} and \gls{DQPSK} in the original 802.11 DSSS (Direct Sequence Spread Spectrum) PHY modes.

Differential-detection-based schemes have also been combined with OFDM systems. This combination enables differential encoding in the subcarrier (frequency) domain instead of in the time domain~\cite{chenhuNonCoherentMassiveMIMOOFDM2020}. This has a lower impact on latency than performing differential encoding between different OFDM symbols, and is also very robust to phase noise. The robustness comes from a component of the phase noise, the common phase error, being the same for all subcarriers and, as a result, getting compensated by the differential decoding. Moreover, differential-detection-based schemes have also been used in combination with beamforming~\cite{chenhuNonCoherentMassiveMIMOOFDM2020}, which helps solve the lack of spatial diversity (due to a usually low number of receive antennas) in the downlink. Some works also show the use of differential detection-based schemes along with a \gls{RIS} in OFDM systems~\cite{chen2022non}. They find that, due to the lack of CSI, the power gain in differential \gls{RIS} is $N_{\rm RIS}$, being $N_{\rm RIS}$ the number of \gls{RIS} elements, instead of the $N_{\rm RIS}^2$ from when a coherent \gls{RIS} is pointed to a user.

In contrast, Grassmannian signaling and energy-based schemes, while theoretically appealing, remain at a more exploratory stage in terms of implementation. Although there has been extensive research on the constellation design, not many works have considered the practical aspects and limitations of their implementation in real hardware equipment.

For instance, a link-level performance study has been conducted in \cite{Cabrejas2017intergration}, where Grassmannian signaling is compared under realistic channel assumptions with the diversity transmission modes standardized in \gls{LTE} (in particular Space-Frequency Block Coding and Frequency-Switched Transmit Diversity). This work shows that in high mobility scenarios, even with substantial antenna correlation, Grassmannian signaling surpasses \gls{LTE} diversity transmission modes when using four or more transmit antennas. Moreover, in the high \gls{SNR} regime, it can enhance the link data rate by up to 10\% with two antennas and 15\% with four antennas.

Some implementations of Grassmannian constellations in an \gls{OFDM} system have been proposed in the literature \cite{Ayadi2009OFDM,Fouad2015OFDM}, but the performance of these proposals has not been evaluated in over-the-air experiments. More recently, another OFDM implementation of Grassmannian signaling has been proposed and implemented in real hardware in \cite{fanjul2017experimental}. In this work, Grassmannian signaling was compared with the coherent Alamouti scheme and the differential Alamouti scheme, showing better performance than the coherent approach but performing slightly worse than the differential Alamouti scheme for fast-fading scenarios. Another recent work that has implemented Grassmannian constellations in an OFDM system is \cite{CuevasThesis2024}, where some new constellation designs \cite{Cuevas21WSA,Cuevas24constellationsOnTheSphere} have been evaluated in real scenarios. 

For energy-based schemes, an illustrative example of energy-based noncoherent communication integrated with OFDM is presented in \cite{Fazeli2020Generalized}, where the authors propose a generalized OFDM with index modulation (OFDM-IM) scheme. This work significantly extends traditional OFDM-IM systems by allowing a variable number of active subcarriers, thereby enabling greater flexibility in code design and improving performance metrics such as diversity gain and error resilience. Importantly, since no information is modulated onto the phase of the subcarriers, the detection relies purely on energy observations. This contribution demonstrates the potential of energy-based noncoherent detection in practical multi-carrier systems and highlights key advantages over coherent schemes in scenarios where channel estimation is costly or unreliable.

An alternative approach to \gls{OFDM} for noncoherent communication over frequency-selective channels is modulation on conjugate-reciprocal zeros (MOCZ)~\cite{Walk2019}. In MOCZ, the information is modulated onto the zeros of the transmitted discrete-time baseband signal's $z$-transform, which is invariant to channel impulse response.

\subsection{Impact of DC Components}

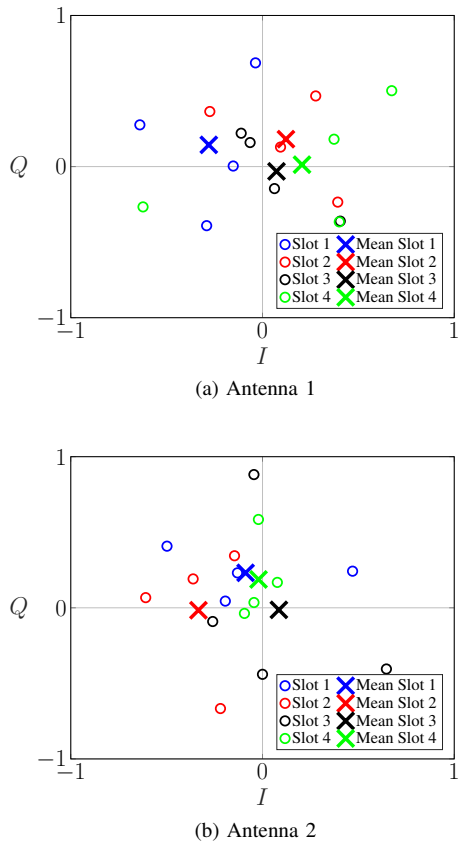
\begin{figure}[t!]
    \centering
    \subfloat[Antenna 1]{
%
%
\definecolor{mycolor1}{rgb}{0.00000,0.44700,0.74100}%
\definecolor{mycolor2}{rgb}{0.85000,0.32500,0.09800}%
\definecolor{mycolor3}{rgb}{0.92900,0.69400,0.12500}%
\definecolor{mycolor4}{rgb}{0.49400,0.18400,0.55600}%
\resizebox{.75\columnwidth}{!}{

\begin{tikzpicture}

\begin{axis}[%
width=4.521in,
height=3.566in,
at={(0.758in,0.481in)},
scale only axis,
font = \huge,
xmin=-1,
xmax=1,
xtick={-1,0,1},
xlabel style={font=\huge\color{white!15!black}},
xlabel={$I$},
ymin=-1,
ymax=1,
ytick={-1,0,1},
ylabel style={font=\huge\color{white!15!black}, rotate = -90},
ylabel={$Q$},
axis background/.style={fill=white},
xmajorgrids,
ymajorgrids,
legend columns = 2,
legend style={font = \Large, at={(0.97,0.03)}, anchor=south east, legend cell align=left, align=left, draw=white!15!black}
]
\addplot [color=mycolor1, line width=1.5pt, only marks, mark size=4.0pt, mark=o, mark options={solid, blue}]
  table[row sep=crcr]{%
-0.152728528092755	0.00372655393005499\\
-0.0366506664116184	0.686345883111596\\
-0.639691330935562	0.276169739878722\\
-0.291342270554663	-0.39093037842241\\
};
\addlegendentry{Slot 1 }

\addplot [color=mycolor1, line width=3.0pt, only marks, mark size=10.0pt, mark=x, mark options={solid, blue}]
  table[row sep=crcr]{%
-0.28010319899865	0.143827949624491\\
};
\addlegendentry{Mean Slot 1}

\addplot [color=mycolor2, line width=1.5pt, only marks, mark size=4.0pt, mark=o, mark options={solid, red}]
  table[row sep=crcr]{%
0.392356174811779	-0.23509753837121\\
-0.274801635207388	0.364387854039711\\
0.093962079569656	0.129223420485031\\
0.277456099516353	0.467465986683075\\
};
\addlegendentry{Slot 2 }

\addplot [color=mycolor2, line width=3.0pt, only marks, mark size=10.0pt, mark=x, mark options={solid, red}]
  table[row sep=crcr]{%
0.1222431796726	0.181494930709152\\
};
\addlegendentry{Mean Slot 2}

\addplot [color=mycolor3, line width=1.5pt, only marks, mark size=4.0pt, mark=o, mark options={solid, black}]
  table[row sep=crcr]{%
-0.111257601579916	0.220792004303417\\
0.0623986464388404	-0.145564410647436\\
-0.0647270173180735	0.159056859381873\\
0.405552925242059	-0.361033391932778\\
};
\addlegendentry{Slot 3 }

\addplot [color=mycolor3, line width=3.0pt, only marks, mark size=10.0pt, mark=x, mark options={solid, black}]
  table[row sep=crcr]{%
0.0729917381957276	-0.0316872347237309\\
};
\addlegendentry{Mean Slot 3}

\addplot [color=mycolor4, line width=1.5pt, only marks, mark size=4.0pt, mark=o, mark options={solid, green}]
  table[row sep=crcr]{%
0.673885939651759	0.502191463793765\\
0.400375086719899	-0.36593552944819\\
-0.623164202501302	-0.266787212092147\\
0.373053374618363	0.18110681702556\\
};
\addlegendentry{Slot 4 }

\addplot [color=mycolor4, line width=3.0pt, only marks, mark size=10.0pt, mark=x, mark options={solid, green}]
  table[row sep=crcr]{%
0.206037549622179	0.012643884819747\\
};
\addlegendentry{Mean Slot 4}

\end{axis}

\begin{axis}[%
width=5.833in,
height=4.375in,
at={(0in,0in)},
scale only axis,
xmin=0,
xmax=1,
ymin=0,
ymax=1,
axis line style={draw=none},
ticks=none,
axis x line*=bottom,
axis y line*=left
]
\end{axis}
\end{tikzpicture}%

}}
    \hfill
    \subfloat[Antenna 2]{
%
%
\definecolor{mycolor1}{rgb}{0.00000,0.44700,0.74100}%
\definecolor{mycolor2}{rgb}{0.85000,0.32500,0.09800}%
\definecolor{mycolor3}{rgb}{0.92900,0.69400,0.12500}%
\definecolor{mycolor4}{rgb}{0.49400,0.18400,0.55600}%
\resizebox{.75\columnwidth}{!}{

\begin{tikzpicture}

\begin{axis}[%
width=4.521in,
height=3.566in,
at={(0.758in,0.481in)},
scale only axis,
font = \huge,
xmin=-1,
xmax=1,
xtick={-1,  0,  1},
xlabel style={font=\huge\color{white!15!black}},
xlabel={$I$},
ymin=-1,
ymax=1,
ytick={-1,  0,  1},
ylabel style={font=\huge\color{white!15!black}, rotate = -90},
ylabel={$Q$},
axis background/.style={fill=white},
xmajorgrids,
ymajorgrids,
legend columns = 2,
legend style={font = \Large, at={(0.97,0.03)}, anchor=south east, legend cell align=left, align=left, draw=white!15!black}
]
\addplot [color=mycolor1, line width=1.5pt, only marks, mark size=4.0pt, mark=o, mark options={solid, blue}]
  table[row sep=crcr]{%
-0.130747189180139	0.231160348841521\\
0.46973391310682	0.242387994501699\\
-0.193992306799775	0.0443460799770941\\
-0.498172555520847	0.407853103269147\\
};
\addlegendentry{Slot 1 }

\addplot [color=mycolor1, line width=3.0pt, only marks, mark size=10.0pt, mark=x, mark options={solid, blue}]
  table[row sep=crcr]{%
-0.0882945345984853	0.231436881647365\\
};
\addlegendentry{Mean Slot 1}

\addplot [color=mycolor2, line width=1.5pt, only marks, mark size=4.0pt, mark=o, mark options={solid, red}]
  table[row sep=crcr]{%
-0.146031517242425	0.344355620069723\\
-0.219355747888735	-0.666678789762629\\
-0.609258662393954	0.0673281426452834\\
-0.36190281778019	0.191168014587851\\
};
\addlegendentry{Slot 2 }

\addplot [color=mycolor2, line width=3.0pt, only marks, mark size=10.0pt, mark=x, mark options={solid, red}]
  table[row sep=crcr]{%
-0.334137186326326	-0.0159567531149429\\
};
\addlegendentry{Mean Slot 2}

\addplot [color=mycolor3, line width=1.5pt, only marks, mark size=4.0pt, mark=o, mark options={solid, black}]
  table[row sep=crcr]{%
-0.0448048654823265	0.881638026034939\\
-4.56768393168732e-05	-0.440188276308778\\
0.646185991904154	-0.404853377352334\\
-0.259770794793679	-0.0907695052852521\\
};
\addlegendentry{Slot 3 }

\addplot [color=mycolor3, line width=3.0pt, only marks, mark size=10.0pt, mark=x, mark options={solid, black}]
  table[row sep=crcr]{%
0.0853911636972078	-0.0135432832278565\\
};
\addlegendentry{Mean Slot 3}

\addplot [color=mycolor4, line width=1.5pt, only marks, mark size=4.0pt, mark=o, mark options={solid, green}]
  table[row sep=crcr]{%
-0.0942542881769979	-0.0372439630249633\\
0.0773354161555508	0.168148954411228\\
-0.0445463194365108	0.0349892231543759\\
-0.0213070324747083	0.584625635095234\\
};
\addlegendentry{Slot 4 }

\addplot [color=mycolor4, line width=3.0pt, only marks, mark size=10.0pt, mark=x, mark options={solid, green}]
  table[row sep=crcr]{%
-0.0206930559831665	0.187629962408969\\
};
\addlegendentry{Mean Slot 4}

\end{axis}

\begin{axis}[%
width=5.833in,
height=4.375in,
at={(0in,0in)},
scale only axis,
xmin=0,
xmax=1,
ymin=0,
ymax=1,
axis line style={draw=none},
ticks=none,
axis x line*=bottom,
axis y line*=left
]
\end{axis}
\end{tikzpicture}%

}}
    \hfill
    \caption{I/Q components of transmitted Grassmannian constellation designed with method proposed in \cite{Cuevas21WSA} for $T = 4$ symbol periods, $M = 2$ antennas and $L = 4$ points from \cite{CuevasThesis2024}. 
    }
    \label{fig:mean_grassmann}
\end{figure}

One problem that arises when trying to implement Grassmannian constellations in practice is that they have non-zero mean, i.e., the I/Q components of the transmitted signal over time do not have zero mean, as is usual in standard coherent modulation formats such as \gls{QAM} or in energy-based or differential detection-based schemes. In other words, coherent schemes do not waste power in transmitting a DC component that carries no information. However, the situation is different with Grassmannian constellations, as can be seen in Fig. \ref{fig:mean_grassmann} showing a scatter plot of the I/Q components of the transmitted signal over 4 consecutive slots for a Grassmannian constellation of size $L = 4$ points designed using the algorithm in \cite{Cuevas21WSA} for $T = 4$ and $M = 2$ transmit antennas. In this figure, we can clearly see that the I/Q components of the transmitted Grassmannian signal over time do not have zero mean. 

In practice, a DC component may cause carrier leakage, baseline wander, and other undesired effects. It is therefore common for front-end receivers to be AC coupled to filter out the DC signal component. If not properly corrected, this DC removal can have a very significant impact on the \gls{SER} performance of noncoherent constellations.

The solution to obtain a zero-mean Grassmannian constellation is simple. The idea exploits the essence of noncoherent communications on the Grassmann manifold: multiplying the transmitted signal by a scalar of unit modulus does not modify the transmitted subspace or the transmitted power. The way to achieve a zero-mean constellation is to transmit $\X s$, where $s$ is a binary random variable taking values in $\{-1,1\}$ with equal probability. 
There is no need in this way to modify the \gls{ML} detector since the constellation has not changed, although now it has a zero-average DC component.

\subsection{Robustness to Hardware Impairments}

Another problem that is typically encountered in practical scenarios is the appearance of \glspl{HWI}, due to which wireless communication devices deviate from their ideal behavior, thus degrading performance \cite{Buzzi2016Survey,Hossain20155G,SoleymaniHWI2020}. These impairments arise from factors such as quantization noise, phase noise, amplifier nonlinearities, I/Q imbalance and frequency offsets caused by mismatched local oscillators \cite{Javed2019HWI,Bjornson2013newlook}. 

While the performance of coherent wireless systems under various \glspl{HWI} has been thoroughly investigated (with several mitigation strategies proposed \cite{Bou2016IQ,Mokhtar2013IQ,SoleymaniHWI2020}), studies on noncoherent systems have been largely restricted to differential modulations \cite{Selim2017Performance,Bucher2020HWIs}. For instance, the study in \cite{Bucher2020HWIs} provides a detailed investigation into the impact of \glspl{HWI} on noncoherent massive MIMO systems employing differential modulation techniques, particularly \gls{DPSK}. This analysis reveals that non-linearities in low-noise amplifiers (LNAs) and coarse quantization from low-resolution analog-to-digital converters (ADCs) (particularly 1-bit converters) severely degrades differential modulations performance. However, phase noise, while generally more problematic for coherent systems, is shown to have a reduced effect in differential detection-based schemes due to the inherent robustness of phase-difference encoding.

For the case of energy-based schemes, an improved energy detector that considers \glspl{HWI} for accurate spectrum sensing has been proposed in \cite{Leila2019Performance}. This work addresses a critical gap in the literature by modeling aggregate transceiver \glspl{HWI}, including I/Q imbalance, amplifier non-linearities and oscillator phase noise, within the framework of an improved energy detector.

\begin{figure}[t!]
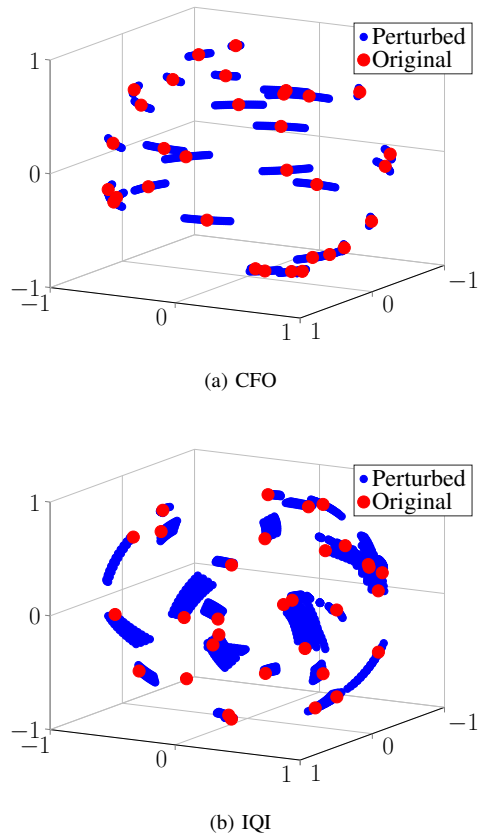

    \centering
    \subfloat[\gls{CFO}]{\input{figs/cfo_grassmann}}
    \hfill
    \subfloat[\gls{IQI}]{\input{figs/iqi_grassmann}}
    \hfill
    \caption{\gls{CFO} and \gls{IQI} effects on the transmitted Grassmannian constellation of $L = 32$ points for $T = 2$ symbol periods and $M = 1$ antenna from \cite{CuevasThesis2024}. (a) \gls{CFO} resulting in a maximum offset of 10$^\circ$ between consecutive samples. (b) \gls{IQI} with maximum amplitude shift of 2 dB and maximum phase shift of 7.5$^\circ$.}
    \label{fig:hwi_grassmann}
\end{figure}

For the case of Grassmannian signaling, a solution has been recently developed in \cite{Cuevas24harware} to counteract the performance degradation due to \glspl{HWI} in Grassmannian constellations. The proposal in \cite{Cuevas24harware} is a \gls{HWI}-aware Grassmannian constellation design robust against \gls{IQI} and \gls{CFO}. In the context of noncoherent Grassmannian constellations, \gls{IQI} and \gls{CFO} are particularly critical since even minor \gls{IQI}s or moderate mismatches between transmitter and receiver oscillators can substantially alter the transmitted subspaces. This effect can be seen in Fig. \ref{fig:hwi_grassmann}, where the original Grassmannian constellation points are shown together with the cloud of possible perturbed codewords for different values of the impairments.


In \cite{Cuevas24harware}, it is assumed that the exact values of these impairments are unknown and, therefore, cannot be directly compensated. Instead, an uncertainty set based on their maximum expected values is defined and the objective is to design Grassmannian constellations that remain robust across all potential values within this set. This approach is especially pertinent in \gls{IoT} scenarios, where numerous low-cost, uncalibrated devices rely on noncoherent communication techniques. 

\subsection{Data Rate Considerations}

Data rate remains a critical metric for assessing the efficiency and practicality of communication schemes, particularly in bandwidth-constrained scenarios. In the context of noncoherent communication approaches, energy-based schemes face inherent limitations due to their reliance solely on signal energy, without exploiting the phase of the received signal. As a result, the constellation of possible received symbols is effectively one-dimensional, which restricts the use of high-order constellations and limits the number of bits that can be transmitted per symbol. This dimensionality constraint imposes a fundamental bottleneck on the achievable data rate. 

Despite this limitation, energy-based constellation schemes offer notable advantages in terms of low implementation complexity and robustness under the conditions outlined in Table \ref{tab:assumptions}, such as spatially uncorrelated channels and accurate power control. This characteristic makes them attractive for low-power, low-complexity \gls{IoT} applications operating in relatively static environments, such as indoor surveillance. For scenarios demanding higher and more continuous data rates, differential modulation schemes and Grassmannian constellations represent more suitable alternatives.

In particular, structured Grassmannian constellations offer a compelling solution for noncoherent systems requiring high spectral efficiency, as they strike a favorable balance between performance and detection computational complexity. Unlike unstructured constellations, which are derived through numerical optimization methods and require the use of computationally intensive \gls{ML} detection across the number of points of the constellation, structured constellations introduce algebraic or geometric regularities that greatly simplify detection. Consequently, while unstructured constellations are typically limited to rates below 1.5 bits/antenna/channel use due to their exponential complexity, structured designs become essential for practical implementation at higher rates, particularly beyond 2 bits/antenna/channel use \cite{Cuevas23unionBound}.

\subsection{Scalability with Number of Users}

Scalability is a critical requirement for multi-user communication systems, particularly in scenarios involving massive connectivity. In this subsection, we evaluate how each noncoherent scheme scales with the number of users and the challenges that arise in multi-user settings.

Energy-based schemes face significant limitations in this context due to their inherently one-dimensional constellations, which constrain the ability to distinguish between multiple users sharing the same time and frequency resources. In such cases, users must be jointly multiplexed within a single-dimension composite constellation \cite{manolakosConstellationDesignNoncoherent2014}. This joint constellation must define a unique decoding region for every combination of user symbols, a condition that leads to highly complex optimization problems and poor scalability as the number of users increases~\cite{manolakosCSINotNeeded2014}.

Differential modulation schemes encounter similar scalability issues, largely due to the averaging effect of symbol decoding across receive antennas, which eliminates the possibility of spatial multiplexing. Like energy-based schemes, differential approaches must rely on joint constellations to multiplex users on shared resources. However, because differential detection-based schemes operate in the complex domain and exploit phase information, they offer a two-dimensional constellation space, providing a somewhat improved capacity for user multiplexing. Despite this advantage, the scalability remains limited, as constructing joint constellations that support a large number of users is both computationally complex and practically infeasible~\cite{lopez-moralesConstellationDesignMultiuser2022}. Moreover, both energy-based and differential detection-based schemes suffer from high sensitivity to power imbalances between users, further complicating their use in large-scale multi-user deployments.

Unlike energy-based and differential detection-based schemes, Grassmannian signaling naturally operates in higher-dimensional spaces and can, in principle, support more efficient user separation when properly designed. However, separating users remains challenging, as the receiver must distinguish among subspaces rather than explicit symbols. Multi-user Grassmannian designs typically rely on assigning orthogonal or near-orthogonal subspaces to different users, which inherently limits the number of simultaneously supported users due to the finite dimensionality of the signal space \cite{Schwarz22codebookTraining,Schwarz23approximate}. Furthermore, most existing work focuses on single-user or two-user cases, and efficient joint encoding or decoding strategies for many users remain an open research problem \cite{Alvarez2023constrained,Ngo2022joint_constellation_design}. While the richer structure of the Grassmann manifold provides potential for scalable designs, practical solutions for massive multi-user scenarios have yet to be developed and validated.




\section{Outlook on Emerging Techniques and Research Directions} \label{sec:future_challenges}

Noncoherent communication in MIMO systems 
has emerged as a viable alternative in scenarios where CSI estimation is costly or impractical. 
%
Despite the significant advancements summarized throughout this survey, noncoherent communication systems remain an open and stimulating research area. In previous sections, we have mentioned several open challenges in the development and implementation of noncoherent communication schemes, related to the complexity-performance tradeoff of noncoherent receivers, as well as the model mismatch due to hardware constraints. 
We next provide a further outlook of emerging techniques, applications, and open problems.

\subsection{Emerging Techniques from Machine Learning}

Recently, novel techniques from machine learning have emerged as promising tools to design the physical layer of communication systems. Methods based on deep learning, including deep unfolding and autoencoder-inspired structures, offer practical means to approximate optimal detection rules while maintaining reasonable computational complexity~\cite{OShea2017}. 
Integrating artificial intelligence (AI) and machine learning into noncoherent massive MIMO systems is an emerging challenge with the potential to significantly enhance detection, signal processing, and resource allocation. Traditional noncoherent receivers rely on complex mathematical optimizations, e.g., in high-dimensional Grassmannian manifolds, which become computationally prohibitive as the number of antennas scales. AI-driven approaches, such as deep learning-based detectors, neural network-assisted constellation design, and reinforcement learning for adaptive modulation, can offer low-complexity, data-driven solutions that outperform classical techniques in highly dynamic environments. Moreover, AI-enabled anomaly detection and adversarial machine learning defenses could enhance the robustness of noncoherent systems against jamming and interference in mission-critical applications. 

We list some representative works on noncoherent MIMO system design based on AI/machine learning as follows. The work~\cite{LearningNC} delves into learning-based approaches for end-to-end design of modulation and signal detection within noncoherent MIMO channels. Simulation-driven optimization is considered, showcasing potential outperformance of traditional Grassmannian designs, even in scenarios with extremely short channel coherence times. The work~\cite{DL_NC1} proposes a novel deep-learning approach for noncoherent detection in massive MIMO systems utilizing DPSK, achieving comparable detection performance to traditional methods, even without user power space profile knowledge. The considered setting is contextualized in 5G scenarios.  The authors of~\cite{Xiaotian2021deep_learning} explored autoencoder-based techniques for Grassmannian constellation design, and implemented the generalized likelihood ratio test (GLRT) detector on a convolutional neural network. 
The work~\cite{MU_DL_NC3} presents a deep-learning approach for implementing noncoherent multiple-symbol detection in DPSK multi-user massive MIMO systems. The proposed scheme showcases superior performance compared to existing decision-feedback differential detection (DFDD) and multiple-symbol differential detection (MSDD) schemes. A quantized neural network for noncoherent MIMO detection in sub-teraHertz communications was proposed in~
\cite{NN_NC}, 
leading to high throughput (above 1 Gbps) and energy efficiency.
We summarize the aforementioned studies in Table~\ref{tab:AI_NC}.

\begin{table*}[!ht]
\renewcommand{\arraystretch}{1.1}
\centering
\caption{Literature overview on noncoherent system design based on machine learning}
\label{TableAI_NC}
\begin{tabular}{ |m{3.5cm}|m{.7cm}|m{3cm}|m{9cm}| } 
        \hline
        \textbf{Design} & \textbf{Year} & \textbf{Scenario} & \textbf{Main idea} \\
        \hline \hline
Wang and Koike-Akino~\cite{LearningNC} & 2021 & single-user MIMO & end-to-end learning design of Grassmannian modulation and signal detection \\
\hline
Mahmoud and El-Mahdy~\cite{DL_NC1} & 2021 & single-user massive MIMO &  a deep-learning-based differential detection scheme without user power space profile knowledge \\
\hline
Xu and Ruyet~\cite{Xiaotian2021deep_learning} & 2021 & single-user MIMO & an autoencoder-based Grassmannian constellation design and an implementation of the generalized likelihood ratio detector on convolutional neural networks \\
\hline
Mahmoud \textit{et al.}~\cite{MU_DL_NC3} & 2021 & multi-user massive MIMO & deep-learning approach for noncoherent multi-user detection with DPSK  \\
\hline
Falempin \textit{et al.}~\cite{NN_NC} & 2022 & single-user MIMO & quantized neural networks (QNN) for noncoherent detection in sub-teraHertz communications \\
\hline
\end{tabular}
\label{tab:AI_NC}
\end{table*}

Despite recent development of machine learning-based solutions, challenges such as real-time training, generalization across diverse channel conditions, and the explainability of AI models in safety-critical systems must be addressed before AI-driven noncoherent massive MIMO solutions can be deployed in practical networks. 

\subsection{Emerging Scenarios and Research Directions}
Noncoherent communication is being used as a design component of emerging application scenarios. We discuss a (non-exhaustive) list of such scenarios in the following. 

    \subsubsection{Massive Random Access} 
    In massive random access, the users' random activity prevents the pre-assignment of orthogonal pilots, making channel estimation challenging. Therefore, noncoherent communication is a potential candidate. Energy-based detection techniques have been used for active user identification in~\cite{Robin23}. In \gls{UMA}~\cite{Liva24_UMA}, where all users accessing the medium share the same codebook and the receiver aims to recover the set of transmitted messages without identifying their senders, 
    Grassmannian constellations have been used as a component to construct the common \gls{UMA} codebook following a tensor-based modulation approach~\cite{Decurninge21tensor,Luan22tensor}.
        
    \subsubsection{Reconfigurable Intelligent Surfaces (RIS)} 
    \gls{RIS} offers an innovative way to control the propagation environment, dynamically modifying channel conditions. 
    \gls{RIS} can provide an extra communication path, especially if the direct channel between the transmitter and the receiver is weak or blocked.
    A challenge in the implementation of \gls{RIS} in the noncoherent setting is that the CSI is not available to control the phase of the \gls{RIS}'s elements. However, in this case, instead of co-phasing with the channel, the \gls{RIS} can send extra information in the phase rotation matrix without affecting the noncoherent detection at the receiver, resulting in extra \gls{DoF} compared to the direct noncoherent link~\cite{Seddik22DoF_RIS}. Although the lack of CSI prevents achieving the full power gain that is quadratic on the number of \gls{RIS} elements~\cite{manuelOJCS_diffOFDM}, this gain can still be achieved if measurements of the received signal power are exploited to point the \gls{RIS} towards the user~\cite{Ren23,Xu24}. In~\cite{manuelOJCS_diffOFDM}, a zero-overhead beam training for \gls{RIS} is proposed, relying on data transmission and reception based on DPSK, that shows similar gains. 
    Therefore, CSI-free \gls{RIS} deployments still have potential and require further investigation.
        
     \subsubsection{Integrated Sensing and Communication (ISAC)}
     ISAC systems are gaining attention for applications such as autonomous vehicles and surveillance radar. By enabling the sharing of spectrum, waveform, or hardware resources between communications and sensing systems, ISAC becomes one of the key technologies of the next generation of wireless networks. Most ISAC systems currently under study consider coherent communications. However, sensing problems that do not require estimating the channel parameters (e.g., delay spread or Doppler shifts) between the target and the receiver can be naturally integrated with noncoherent communication schemes. One recent result that motivates this line of work is the characterization of the Pareto frontier of the sensing and communication performances in \cite{Xiong23tradeoff}. This work proves that, asymptotically, the optimal sensing strategy that minimizes the Cramér-Rao lower bound (CRLB), constrained by a maximum communication rate, is achieved by sending waveforms uniformly distributed on the Stiefel manifold. Although the communications over the Stiefel manifold are inherently coherent since it is necessary to distinguish two distinct bases of the same subspace, this asymptotic result suggests the interest of studying waveforms specifically optimized on the Grassmannian for noncoherent communications. Another recent work that uses unitary constellations as ISAC waveforms is \cite{FodorTCOM25}. There is a potential to apply manifold-constrained optimization techniques used for noncoherent constellation design to the problem of ISAC waveform design. 
        
    \subsubsection{New Waveform, e.g., Orthogonal Time Frequency Space (OTFS)} Traditional waveforms such as OFDM are not optimized for fast varying channels. 
    Alternatives like \gls{OTFS}, designed for high-Doppler environments, could be integrated with noncoherent schemes to improve communications over doubly-selective channels when channel estimation is challenging. It has recently been shown that coherent OTFS systems, which assume perfect CSI knowledge of the time-invariant channel in the delay-Doppler plane, require an 
    subcarrier spacing 
    greater than twice the maximum Doppler frequency. This limitation of coherent OTFS systems is shown in \cite{LajosOTFSTCOM24}, where a noncoherent OTFS scheme, based on differential detection schemes, is proposed that is capable of operation when the subcarrier spacing is larger than the maximum Doppler frequency (not twice). More work is needed to explore the pros and cons of noncoherent OTFS systems. 
        

    \subsubsection{Low Probability of Detection (LPD) Communications} 
    Another scenario in which noncoherent schemes may have application and require further investigation is that of LPD communication \cite{bash_hiding_2015}. In an LPD system, a trusted transmitter (Alice) attempts to communicate with a trusted receiver (Bob) while a warden (Willie) monitors the channel, trying to detect the communication. Typically, LPD communications are based on coherent signaling, and it is common to find the assumption of perfect CSI in the literature. However, obtaining accurate CSI requires sending many pilots, which would provide reference signals facilitating Willie's detection of the transmission, due to the introduction of periodicity in the frame structure. It is therefore necessary to analyze the performance of noncoherent LPD communications systems \cite{HanlyLPD23}. One work along this line is \cite{katsuki_new_2023}, which proposes a noncoherent LPD scheme based on differentially-encoded symbols that follow a Gaussian distribution to make the transmitted signal as close to the noise as possible. 


    \subsubsection{Physical-Layer Security without CSI}
    Secure communications in noncoherent systems present unique challenges and opportunities. Since noncoherent communication does not rely on explicit CSI, it is inherently more resistant to eavesdropping in dynamic environments. However, there are still open issues related to securing the physical layer. 
            Traditional physical layer security methods rely on beamforming and secret key generation based on CSI knowledge, which is unavailable in noncoherent systems. Alternative security strategies, such as artificial noise generation in the Grassmannian domain or secure modulation techniques that exploit statistical CSI~\cite{Choi21}, must be developed.
            Furthermore, noncoherent massive MIMO systems
could be vulnerable to jamming attacks and spoofing, where an attacker exploits statistical properties of the channel. Robust signal processing and adversarial machine learning techniques could improve resilience.
            Finally, the lack of CSI opens the door to novel keyless authentication methods, leveraging device fingerprints, energy-based modulation variations, or subspace-based user identification.

\section{Conclusions} \label{sec:conclusion}
\note{Hoang}

This survey has demonstrated that noncoherent communication is a compelling and increasingly necessary paradigm for future wireless systems, particularly in scenarios where acquiring accurate CSI is costly, unreliable, or infeasible. By reviewing the theoretical foundation and design methodologies of three principal approaches\textemdash Grassmannian signaling, differential detection, and energy-based schemes\textemdash we showed how each exploits distinct structural invariants of the channel to enable signal recovery without explicit channel estimation. 
We also highlighted important implementation challenges, such as compatibility with OFDM, resilience to hardware impairments, and scalability with the number of users, which must
be addressed when deploying noncoherent schemes in practical wireless communication systems.
Finally, we presented an outlook on the role of noncoherent communication in emerging applications and open challenges that need to be resolved for noncoherent communication
to be adopted in next-generation wireless systems.

\begin{appendices}

\section{Background on Riemannian manifolds} \label{app:manifold}


We present a brief introduction to Riemmanian geometry, focusing on the complex Grassmann manifold. A more detailed account of Riemannian manifolds and, in particular, Grassmann and Stiefel manifolds can be found in \cite{AbsMahSep2008optManifolds, Edelman1999GAO} \cite[Chap. 9]{Coherence}.

\subsection{Grassmann and Stiefel Manifolds}
The complex Grassmannian $\Gras$ is the set of $M$--dimensional complex subspaces of $\CC^T$, with $T>M$, that is a complex manifold of dimension $M(T-M)$. Elements in $\Gras$ are represented by matrices in the Stiefel manifold $\Am\in\St$, that is $\Am\in\CC^{T\times M}$, $\Am^\H\Am=\Id_\Nt$. This representation is not unique, since $\Am$ and $\Am\Um$ with $\Um$ a unitary $M\times M$ matrix represent the same element in $\Gras$, so formally we denote elements of the Grassmannian as $[\Am]$ where $\Am\in\St$ and $[\Am]$ is the class of $\Am$ under the quotient by the set of $M\times M$ unitary matrices $\mathcal{U}_M$.

\subsection{Subspace Distances}
To measure the distance between two subspaces, we need the concept of principal angles \cite{GoluVanl96matrix_computations},\cite{Golub73}. Let $[\Am]$ and $[\Bm]$ be $M$-dimensional subspaces of $\CC^T$ and suppose that $\Am$ and $\Bm$ are two matrices whose columns form orthonormal bases for $[\Am]$  and $[\Bm]$. The smallest principal angle $\theta_1$ is the minimum angle formed by a pair of unit vectors $(\uv_1, \vv_1)$ drawn from $([\Am], [\Bm])$. The second principal angle $\theta_2$ is defined as the smallest angle attained by a pair of unit vectors $(\uv_2, \vv_2)$ that is orthogonal to the first pair, and so on. The sequence of principal angles is non-decreasing, and it is contained in the range $\theta_i \in [0,\pi/2]$.

A more computational definition of the principal angles is presented in \cite{Golub73}: the singular values of the matrix $\Cm = \Am^\H \Bm$ are the cosines of the principal angles between $[\Am]$ and $[\Bm]$: $\cos \theta_1,\ldots, \cos \theta_M$. In other words, the $M \times M$ matrix $\Cm = \Am^\H \Bm$ has the singular value decomposition $\Am^\H \Bm = \Um \Dm \Vm^\H$, where $\Dm = \diag(\cos \theta_1,\ldots, \cos \theta_M)$, and $\Um$  ad $\Vm$ are $M\times M$ unitary matrices. This definition of the principal angles is convenient numerically because singular value decompositions can be computed efficiently with standard linear algebra software packages. Note also that this definition of the principal angles does not depend on the choice of bases that represent the two subspaces.

The principal angles induce several distance metrics, which can be used in Grassmannian packing problems. Recall that computations on Grassmann manifolds are performed using unitary matrix representatives for the points, so any measure of distance must be unitarily invariant. The following are the most widely used \cite{Edelman1999GAO, Dhillon2008constructing}:

\begin{itemize}\setlength{\itemsep}{.3cm}
\item \textbf{Geodesic distance}: the geodesic distance is defined as
\begin{equation}
\label{eq:geodistance}
\mathrm{d}_{\mathrm{geo}} \left( [\Am], [\Bm]  \right) = \left( \sum_{m=1}^{M} \theta_m^2 \right)^{1/2}\,.
\end{equation} 

This distance takes values between zero and $\sqrt{M} \pi/2$. It measures the geodesic distance between two subspaces on the Grassmann manifold. This distance function has the drawback of not being differentiable everywhere. As another drawback, there is no way to isometrically embed $\Gras$ into some Euclidean space in such a way that the geodesic distance \eqref{eq:geodistance} is the Euclidean distance in that space \cite{Conway1996packing}.

\item \textbf{Chordal distance}: the Grassmannian $\Gras$ can be embedded into a Euclidean space of dimension $M(T-M)$, or higher, such that the distance between subspaces is represented by the distance in that space. Some of these embeddings map the subspaces to points on a sphere so that the straight-line distance between points on the sphere is the chord between them, which naturally explains the name {\it chordal distance} for such a metric. Different embeddings are possible and therefore one may find different ``chordal'' distances defined in the literature. However, the most popular embedding that defines a chordal distance is given by the projection matrices. The standard chordal distance between two subspaces is given by 
\begin{align}
\mathrm{d}_{\mathrm{ch}} \left( [\Am], [\Bm]\right) &= \left(\sum_{m=1}^{M} \sin^2 \theta_m \right)^{1/2} \nonumber \\ &= \left(M-\left\|\Am^\H\Bm \right\|^2_{\rm F} \right)^{1/2} \nonumber \\ &= \sqrt{M-\trace(\Bm^\H \Am \Am^\H \Bm)}\,.
\label{eq:FrobNorm}
\end{align}

This is the metric referred to as chordal distance in the majority of works regarding the design of Grassmannian constellations \cite{Conway1996packing,Dhillon2008constructing,BekoTSP2007noncoherentColoredNoise,Gohary2009GrassmanianConstellations}, although it might well be called projection distance or projection F-norm, as in \cite{Edelman1999GAO}, or simply extrinsic distance as in \cite{Srivastava02}. Note that the chordal distance between two subspaces is also given by the Frobenius norm of the difference between the respective projection matrices as
\begin{equation}
\mathrm{d}_{\mathrm{ch}} \left( [\Am], [\Bm]\right) = \frac{1}{\sqrt{2}} \| \Am \Am^\H -  \Bm \Bm^\H\|_{\rm F}.
\label{eq:DchordalProj}
\end{equation}

\item \textbf{Procrustes distance}: the Procrustres distance, frequently used in shape analysis \cite[Chapter 9]{Chikusebook}, is given by

\begin{align}
    \mathrm{d}_{\mathrm{Proc}} \left([\Am],[\Bm]\right) &= \left(\sum_{m=1}^{M} \sin^2 \left(\frac{\theta_m}{2}\right)\right)^{1/2} \nonumber \\ &= \left(M-\left\|\Am^\H\Bm \right\|_*\right)^{1/2}\,,
\end{align}

\noindent where $\left\|\Am^\H\Bm \right\|_*$ denotes the nuclear norm of $\Am^\H\Bm$. The Procrustes distance for the Grassmannian is defined as the smallest Euclidean distance between any pair of matrices in the two corresponding equivalence classes.

\item \textbf{Spectral distance}: the spectral distance is defined as
\begin{equation}
\mathrm{d}_{\mathrm{spec}} \left( [\Am], [\Bm]  \right) = \sin \theta_1 
= \sqrt{1 - \left\|\Am^\H\Bm \right\|_2^2} \,,
\end{equation}

\noindent where $\theta_1$ is the smallest principal angle and $\left\|\Am^\H\Bm \right\|_2$ denotes the $l_2$ or spectral norm of $\Am^\H\Bm$. It takes values between 0 and 1. 

\item \textbf{Fubini-Study distance}: the Fubini-Study distance is given by
\begin{align}
\mathrm{d}_{\mathrm{FS}} \left( [\Am], [\Bm]  \right) &= \text{arccos}\left(\prod_{m=1}^M \cos(\theta_m)\right) \nonumber \\
&= \text{arccos}\left(\det(\Am^\H\Bm)\right) \,,
\end{align}

\noindent which takes values between 0 and $\pi/2$.
\end{itemize}

\subsection{Packing Limits}
The problem of finding packings in the Grassmann manifold with optimal packing efficiencies is not only of interest from a communications point of view but also in some other fields such as compressed sensing, digital fingerprinting, quantum state tomography \cite{QSTArxiv2020}, and multiple description coding. Some known bounds for the minimum chordal distance between two Grassmannian points of a packing, denoted by $d_{\rm ch}$, are presented below: 

\begin{itemize}\setlength{\itemsep}{.3cm}
    \item \textbf{Welch-Rankin bound}: this upper bound was first proved on spherical packings in \cite{Rankin_1955} and is given by
    \begin{equation}
        \mathrm{d}_{\mathrm{ch}} \leq \sqrt{\frac{M(T-M)}{T} \frac{L}{L-1}} \hspace{0.3cm},\hspace{0.3cm} L > T\,.
        \label{eq:rankin_bound}
    \end{equation}

    \item \textbf{Orthoplex bound}: this bound was obtained in \cite{Conway1996packing} and it states
    \begin{equation}
        \mathrm{d}_{\mathrm{ch}} \leq \sqrt{\frac{M(T-M)}{T}} \hspace{0.3cm},\hspace{0.3cm} L > T^2\,.
        \label{eq:orthoplex_bound}
    \end{equation}

    \item \textbf{Levenstein bound}: the original proof of this bound appears in \cite{levenstein} and is given for $M = 1$ in closed form by
    \begin{equation}
        \mathrm{d}_{\mathrm{ch}} \leq \sqrt{\frac{L(T-1)}{(L-T)(T+1)}} \hspace{0.3cm},\hspace{0.3cm} L > T^2\,.
        \label{eq:levenstein_bound}
    \end{equation}
    The approach used to obtain this bound is based on linear programming methods that exploit the harmonic analysis on the underlying manifold. The main idea is to construct an appropriate polynomial (or family of polynomials) which is nonnegative on the spectrum of the distance function and then optimize over it. This yields an upper bound on the best-possible minimum distance. Although the explicit closed form is more complicated than for the previous bounds, the Levenstein bound often provides the sharpest known results in many regimes.
    
\end{itemize}

Some other packing efficiency limits are provided in \cite{Barg2002BoundsOP,Henkel2005spherePackingBounds,Dai2008quantizationBounds,Ngo_cubesplit_journal}. The Gilbert-Varshamov and Hamming bounds for packings of spheres in the Grassmann manifold are proposed in \cite{Barg2002BoundsOP}. In \cite{Henkel2005spherePackingBounds}, some bounds are derived by applying the Riemannian geometry of volume estimates in terms of curvature. The authors in \cite{Dai2008quantizationBounds} computed a closed-form expression of a metric ball of radius at most one in the Grassmannian that they use to derive several bounds. \cite[Lemma~1]{Ngo_cubesplit_journal} gives both upper and lower bounds on the minimum chordal distance for the case $M=1$.

\section{Optimized Packings} \label{app:best_packings}

In this appendix, we present the minimum chordal distances (rounded to the sixth digit) for different packings in $\Gras$ for various values of $T$ (dimension of the ambient space), $M$ (subspace dimension), and $L$ (number of points in the Grassmannian). These results complement and extend the best packings in the complex projective space ${\mathbb{G}}(1,{\mathbb{C}}^{T})$ that can be found in \cite{jasper2019game} and \cite{Park22Allerton} (see also \href{https://www.math.colostate.edu/~king/GameofSloanes.html#JKM19}{Game of Sloanes}). 
The packings have been numerically optimized with the optimization algorithm described in \cite{Cuevas21WSA} and are available at \href{https://github.com/diegocuevasfdez/grassbox}{https://github.com/diegocuevasfdez/grassbox}.\footnote{It should be noted that the presented packings are not claimed to be optimal. Should improved packings with higher minimum chordal distances become available, we welcome contributions and will update the corresponding results in the designated repository accordingly.}

The packings shown in Table \ref{tab:PackingsM1} can be useful as non-orthogonal pilot sequences to mitigate pilot contamination in cell-free massive MIMO (CF-mMIMO) systems, while the packings in Table \ref{tab:PackingsM2} are optimized Grassmannian constellations useful as a benchmark in noncoherent communication problems. 


\begin{table}[H]
\renewcommand{\arraystretch}{1.1}
\caption{Minimum chordal distance for numerically optimized Grassmannian packings in ${\mathbb{G}}(1,{\mathbb{C}}^{T})$}
\centering
\begin{tabular}{ |c|l||c| } 
\hline
$T, M$ & $L$ & $\min$ $\mathrm{d}_\mathrm{c}$ \\
\hline \hline
\multirow{6}{*}{$T=16$, $M=1$} 
& $L=32$ & 0,983454 \\ \cline{2-3}
& $L=64$ & 0,968907 \\ \cline{2-3}
& $L=128$ & 0,955749 \\ \cline{2-3}
& $L=256$ & 0,937461\\ \cline{2-3}
& $L=512$ & 0,918361\\ \cline{2-3}
& $L=1024$ & 0,900135\\ \cline{2-3}

\hline\hline
\multirow{5}{*}{$T=32$, $M=1$} 
& $L=64$ & 0,991442\\ \cline{2-3}
& $L=128$ & 0.985761\\ \cline{2-3}
& $L=256$ & 0,979851\\ \cline{2-3}
& $L=512$ & 0,973092\\ \cline{2-3}
& $L=1024$ & 0,962715\\ \cline{2-3}  
\hline
\end{tabular}
\label{tab:PackingsM1}
\end{table}

\begin{table}[H]
\renewcommand{\arraystretch}{1.1}
\caption{Minimum chordal distance for numerically optimized Grassmannian packings in ${\mathbb{G}}(2,{\mathbb{C}}^{T})$}
\centering
\begin{tabular}{ |c|l||c| } 
\hline
$T, M$ & $L$ & $\min$ $\mathrm{d}_\mathrm{c}$ \\
\hline \hline
\multirow{8}{*}{$T=4$, $M=2$} & $L=8$ & 1,069042 \\ \cline{2-3}
& $L=16$ & 1,032795 \\ \cline{2-3}
& $L=32$ & 0,967701 \\ \cline{2-3}
& $L=64$ & 0,902535 \\ \cline{2-3}
& $L=128$ & 0,830740 \\ \cline{2-3}
& $L=256$ & 0,767332\\ \cline{2-3}
& $L=512$ & 0,657829 \\ \cline{2-3}
& $L=1024$ & 0,625758\\ \cline{2-3}

\hline\hline
\multirow{8}{*}{$T=6$, $M=2$} & $L=8$ &  1,234257\\ \cline{2-3}
& $L=16$ & 1,192500\\ \cline{2-3}
& $L=32$ & 1,160567 \\ \cline{2-3}
& $L=64$ & 1,109148\\ \cline{2-3}
& $L=128$ & 1,069076\\ \cline{2-3}
& $L=256$ & 1,028612\\ \cline{2-3}
& $L=512$ & 0,990052\\ \cline{2-3}
& $L=1024$ & 0,950608\\ \cline{2-3}
\hline
\end{tabular}

\label{tab:PackingsM2}
\end{table}

\end{appendices}

\bibliographystyle{IEEEtran}
\bibliography{references.bib}
\vfill

    \end{document}